\documentclass[aps,prl,preprint,superscriptaddress, notitlepage]{revtex4-1}

\usepackage{graphicx}
\usepackage{gensymb}
\usepackage{booktabs}
\usepackage{array}
\usepackage{bm}

\begin{document}

%\preprint{}

\title{Extreme Ultraviolet Superfluorescence in Xenon and Krypton}



\author{L. Mercadier}
\email{laurent.mercadier@xfel.eu}
\affiliation{Max Planck Institute for the Structure and Dynamics of Matter, 22761 Hamburg, Germany}
\affiliation{European XFEL, 22869 Schenefeld, Germany}
\author{A. Benediktovitch}
\affiliation{Deutsches Elektronen-Synchrotron (DESY), 22761 Hamburg, Germany}
\author{C. Weninger}
\affiliation{Max Planck Institute for the Structure and Dynamics of Matter, 22761 Hamburg, Germany}
\author{M. A. Blessenohl}
\affiliation{Max-Planck-Institut f\"ur Kernphysik, 69117 Heidelberg, Germany}
\author{S. Bernitt}
\affiliation{Institut f\"ur Optik und Quantenelektronik, Friedrich-Schiller-Universit\"at Jena, 07743 Jena, Germany}
\affiliation{Max-Planck-Institut f\"ur Kernphysik, 69117 Heidelberg, Germany}
\author{H. Bekker}
\affiliation{Max-Planck-Institut f\"ur Kernphysik, 69117 Heidelberg, Germany}
\author{S. Dobrodey}
\affiliation{Max-Planck-Institut f\"ur Kernphysik, 69117 Heidelberg, Germany}
\author{A. S\'anchez-Gonz\'alez}
\affiliation{Department of Physics, Imperial College London, London SW7 2AZ, United Kingdom}
\author{B. Erk}
\affiliation{Deutsches Elektronen-Synchrotron (DESY), 22761 Hamburg, Germany}
\author{C. Bomme}
\affiliation{Deutsches Elektronen-Synchrotron (DESY), 22761 Hamburg, Germany}
\author{R. Boll}
\affiliation{Deutsches Elektronen-Synchrotron (DESY), 22761 Hamburg, Germany}
\author{Z. Yin}
\affiliation{Deutsches Elektronen-Synchrotron (DESY), 22761 Hamburg, Germany}
\affiliation{Max Planck f\"ur biophysikalische Chemie, 37077 Göttingen, Germany}
\author{V. P. Majety}
\affiliation{Max Planck Institute for the Structure and Dynamics of Matter, 22761 Hamburg, Germany}
\author{R. Steinbr\"ugge}
\affiliation{Max-Planck-Institut f\"ur Kernphysik, 69117 Heidelberg, Germany}
\author{M. A. Khalal}
\affiliation{Laboratoire de Chimie Physique - Mati\`ere et Rayonnement, Universit\'e Pierre et Marie Curie, F-75231 Paris Cedex 05, France}
\author{F. Penent}
\affiliation{Laboratoire de Chimie Physique - Mati\`ere et Rayonnement, Universit\'e Pierre et Marie Curie, F-75231 Paris Cedex 05, France}
\author{J. Palaudoux}
\affiliation{Laboratoire de Chimie Physique - Mati\`ere et Rayonnement, Universit\'e Pierre et Marie Curie, F-75231 Paris Cedex 05, France}
\author{P. Lablanquie}
\affiliation{Laboratoire de Chimie Physique - Mati\`ere et Rayonnement, Universit\'e Pierre et Marie Curie, F-75231 Paris Cedex 05, France}
\author{A. Rudenko}
\affiliation{J. R. Macdonald Laboratory, Department of Physics, Kansas State University, Manhattan, KS 66506, USA}
\author{D. Rolles}
\affiliation{Deutsches Elektronen-Synchrotron (DESY), 22761 Hamburg, Germany}
\affiliation{J. R. Macdonald Laboratory, Department of Physics, Kansas State University, Manhattan, KS 66506, USA}
\author{J. R. Crespo L\'opez-Urrutia}
\affiliation{Max-Planck-Institut f\"ur Kernphysik, 69117 Heidelberg, Germany}
\author{N. Rohringer}
\email{nina.rohringer@desy.de}
\affiliation{Max Planck Institute for the Structure and Dynamics of Matter, 22761 Hamburg, Germany}
\affiliation{Deutsches Elektronen-Synchrotron (DESY), 22761 Hamburg, Germany}
\affiliation{Department of Physics, Universit\"at Hamburg, 20355 Hamburg, Germany}
%\thanks{}

\date{\today}

\begin{abstract}
We present a comprehensive experimental and theoretical study on superfluorescence in the extreme ultraviolet wavelength regime. Focusing a high-intensity free-electron laser pulse in a cell filled with Xe or Kr gas, the medium is quasi instantaneously population-inverted by inner-shell ionization on the giant resonance followed by Auger decay. On the timescale of 100 ps a macroscopic polarization builds up in the medium, resulting in superfluorescent emission of several Xe and Kr lines in the forward direction. As the number of emitters in the system is increased by either raising the pressure or the pump-pulse energy, the emission shows an exponential growth of over 4 orders of magnitude and reaches saturation. With increasing yield, we observe line broadening, a manifestation of superfluorescence in the spectral domain. Our novel theoretical approach, based on a full quantum treatment of the atomic system and the irradiated field, shows quantitative agreement with the experiment and supports our interpretation.
\end{abstract}

%\pacs{}
%\keywords{}

\maketitle

\begin{figure}[!htb]%
\includegraphics[width=8.6cm]{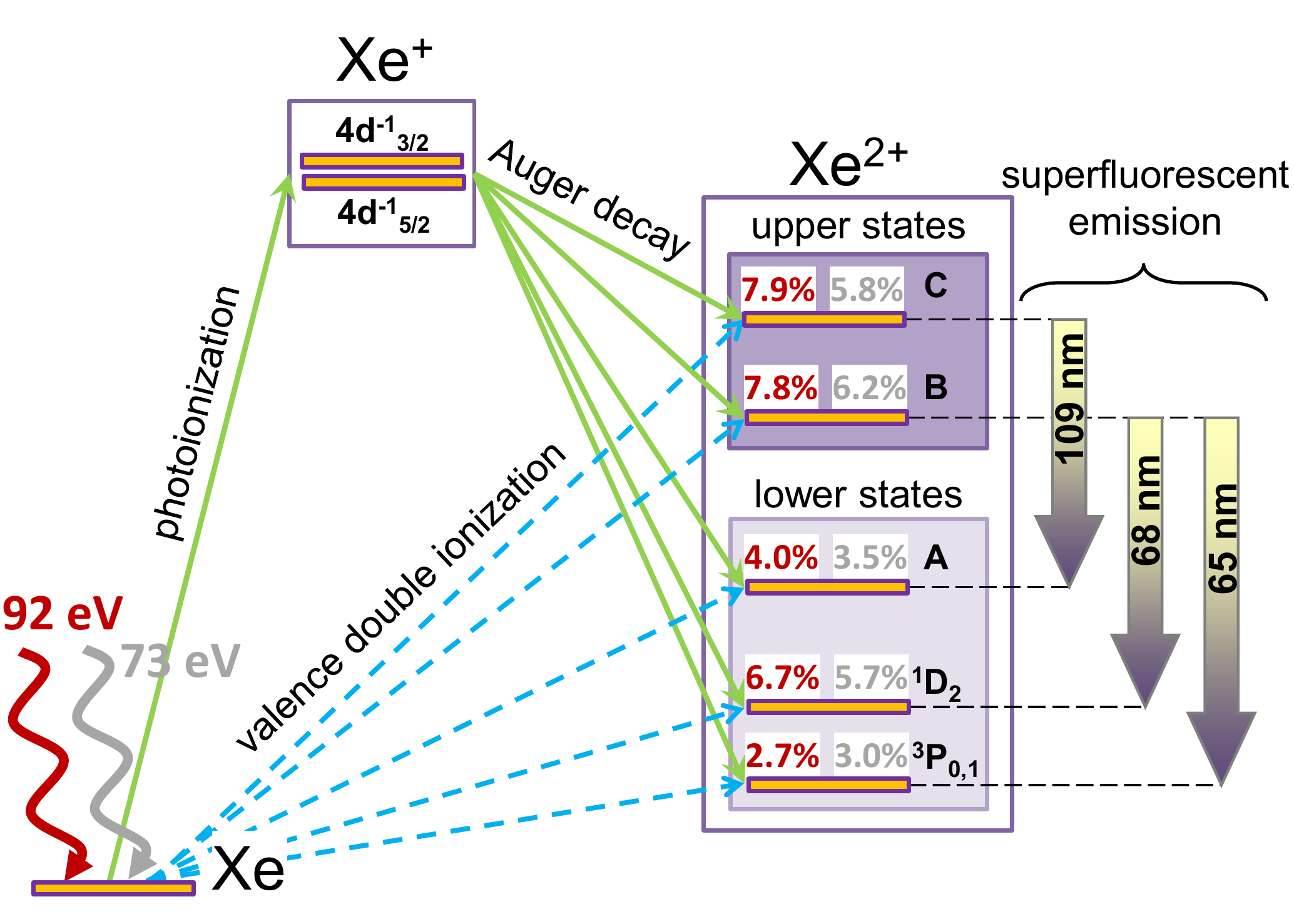}%
\caption{Level scheme for Xe: The FEL pulse photoionizes the $4d$ shell of the Xe ground state. The resulting Xe$^+$ $4d^{-1}$ vacancies decay via Auger process into various Xe$^{2+}$ and Xe$^{3+}$ (not shown) states. In the residual Xe$^{2+}$, population inversion is established: States $B$ and $C$ serve as the upper states and states $A$, $^1D_2$ and $^3P_1$ are the lower states of the observed transitions. Direct double ionization (dashed blue arrows) also contributes to the population of the Xe$^{2+}$ states. The population branching ratios of each state, relative to all Xe$^+$, Xe$^{2+}$ and Xe$^{3+}$ states are deduced from our coincidence measurements and indicated in red (gray) for 92 eV (73 eV) incident photon energy.}
\label{fig:Xe_scheme}
\end{figure}

Superfluorescence \cite{GROSS1982301} is the spontaneous, collective decay of an extended ensemble of atoms that have been prepared in a population-inverted state, resulting in collimated, high-intensity radiation pulses. The pulses are emitted at a certain delay following excitation and have a duration that can be several orders of magnitude smaller than the typical upper-state lifetimes. Long before the advent of short-wavelength free-electron lasers (FELs), strong superfluorescence in optically thick media was proposed as a source of highly intense and pulsed extreme-ultraviolet (XUV) or X-ray radiation \cite{PhysRevLett.35.844}. Strong X-ray K-$\alpha$ superfluorescence following ionization of the $1s$ shell with a focused X-ray FEL (XFEL) beam was demonstrated in neon gas \cite{RohringerNat12,WeningerPRL13}, solid copper \cite{YonedaNat15} and manganese salts in aqueous solution \cite{PhysRevLett.120.133203}. Extremely high gains were observed in these experiments  \cite{RohringerNat12,PhysRevLett.120.133203}, with exponential amplification factors surpassing 20 compared to spontaneous emission. In the vacuum ultraviolet and XUV regions, superfluorescence following inner-shell ionization has so far not been demonstrated. The difficulty to obtain transient gain in this wavelength regime is a consequence of the very different time scales of two competing processes: on the one hand, short (fs) Auger lifetimes of inner-valence vacancies, and, on the other hand, comparatively long (ns) radiative transition times –- a highly unfavorable combination to sustain a sizable population inversion and gain. Here, we present combined experimental and theoretical work, giving strong evidence for XUV superfluorescence of Xe and Kr gases. Population inversion is achieved by photoionization of an inner-shell with an XUV FEL, followed by rapid Auger decay of the inner-shell vacancies. Different Auger-decay channels result in the occupation of excited dicationic states and the creation of a population inversion (see level scheme in Fig. \ref{fig:Xe_scheme}), that sustains coherent collective emission. These Auger-decay pumped lasers, predicted in 1975 \cite{PhysRevLett.35.844}, have been experimentally realized in the 1980s in Xe and Kr gas \cite{KapteynPRL86,KapteynPRA88} using a laser-generated plasma XUV source. Already then, it was speculated that the emission could stem from superfluorescence. The transverse pump geometry and the long rise time of the plasma-generated emission, however, were not ideal to sustain superfluorescent emission. FELs offer a clear advantage: Firstly, their ultra-short pulses at high intensities guarantee a nearly instantaneous inversion of the level population in an extended ensemble of atoms. Secondly, the transverse coherence of FELs allows for tight beam focusing and the possibility to longitudinally pump a long and narrow gain medium in a so-called gain-swept pumping geometry \cite{PhysRevA.11.989}, which is ideal for creating superfluorescent emission \cite{Sete_IEEE12}. As shown in Refs. \cite{PhysRevLett.107.193603,doi:10.7566/JPSJ.84.054301, doi:10.7566/JPSJ.85.034301}, FEL excitation of He gas induces optical superfluorescence. Here, we report superfluorescence in the XUV spectral region. 
\\
%%%%%%%%%%%%%%%%%%%%%%%%%%%%%%%%%%%%%%%%%%%%%%%%%%%%%%
%%%%%%%%%%%%%%%%%%%%%%%%%%%%%%
Superfluorescence is characterized in the temporal domain by a highly directional emission peaking around a characteristic delay $\tau_D$ following the excitation, and is often accompanied by temporal ringing. The pulse peak power grows quadratically with the number of emitters. Another important feature, the manifestation of superfluorescence in the spectral domain, was hardly studied in the past \cite{book'superradiance,1982'Malikov,MALIKOV198474}. Here, we present a comprehensive spectroscopic measurement of the superfluorescence in Xe and Kr. By increasing the number of emitters as we raised the pump-pulse energy and pressure, we can make a quantitative comparison to our novel theory \cite{andrei}. Our theoretical approach is fully quantized in both atomic and field degrees of freedom and goes beyond the typical Maxwell-Bloch-like phenomenological treatments \cite{GROSS1982301,WeningerPRL13,WeningerPRA14}. It predicts ensemble averages of the temporal and spectral intensity profiles of the emission, and enables a realistic treatment of the pump process.

%%%%%%%%%%%%%%%%%%%%%%%%%%%%%%
The experiment was performed at the CAMP end station of the Free-Electron Laser in Hamburg (FLASH) \cite{Erk_CAMP}. Pulses of $\sim80 - 100$ fs duration at 10-Hz repetition rate were focused into a pressurized gas cell \cite{RohringerNat12, WeningerPRL13} to result in an effectively pencil-shaped pumped medium 
of 4.5 mm length and 20-30~$\mu$m radius. The photon energy was tuned to 73~eV, 92~eV and 100~eV (across the giant $4d$ resonance that peaks at around 100 eV \cite{Becker_PRA89}) in Xe, and to 100~eV for Kr. The maximum available pulse energy on target was 90~$\mu$J. The transmitted FEL pulse and the XUV line emission were analyzed by a high-resolution spectrometer allowing the measurement of both the FEL spectrum and the lasing lines in different diffraction orders. Given the small solid angle of detection of $1.4\times10^{-4}$~sr, our setup did not measure the mere fluorescence emitted by the excited gas in the forward direction.. 

\begin{figure}[!htb]%
\includegraphics[width=8.6cm]{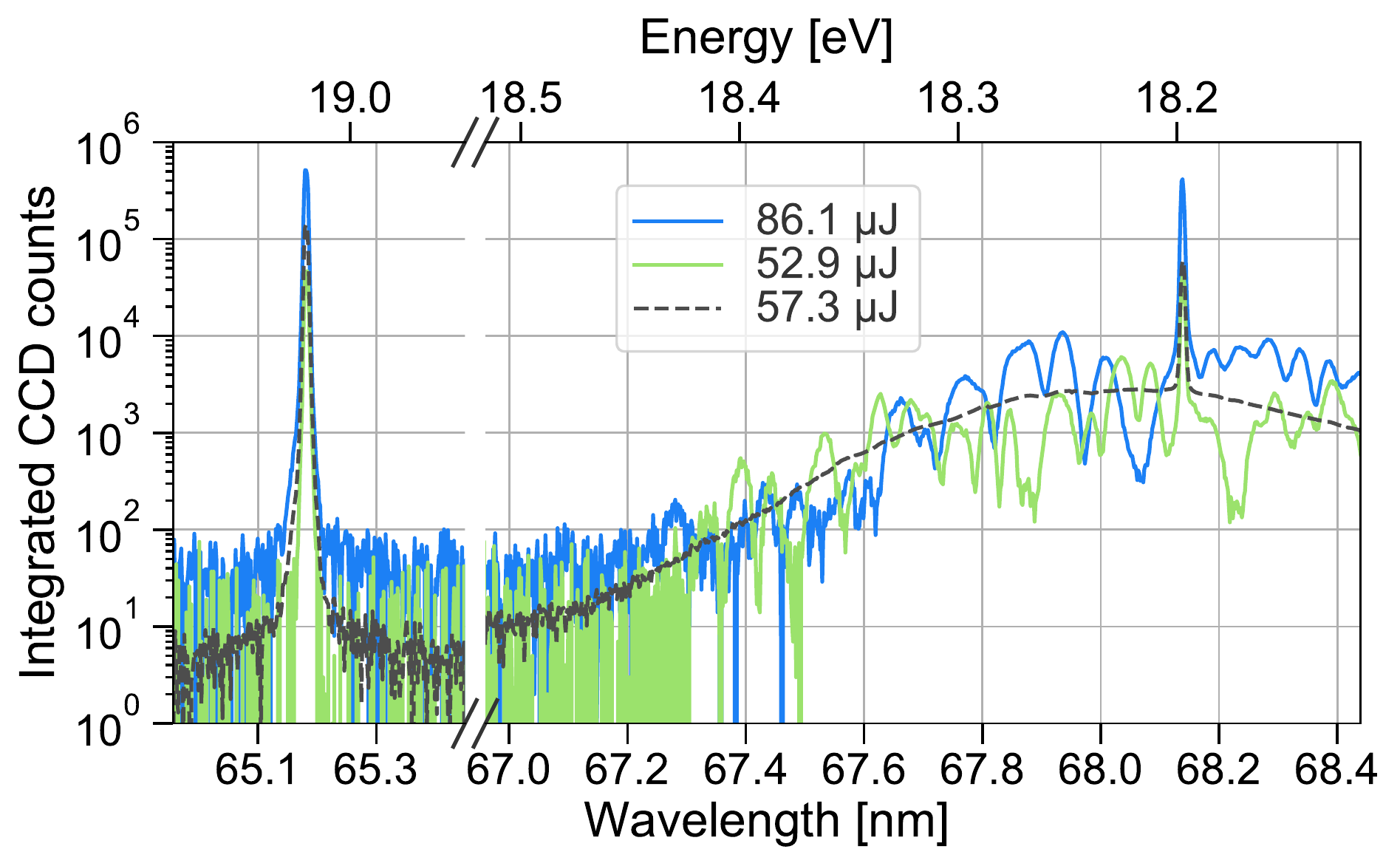}%
\caption{Single-shot spectra (solid blue and green) and average over 725 spectra (dashed black) obtained with an FEL photon energy of 73 eV in 7 mbar of Xe. Two laser-like emission lines at 65.18 nm and 68.14 nm are observed. The spectrum of a single transmitted FEL pulse appears in 4\textsuperscript{th} diffraction order around 18.2 eV. It shows randomly distributed spikes contained in a $\sim 1$~eV bandwidth from self-amplified stimulated emission. An averaged spectrum that includes the shot-to-shot central wavelength fluctuations has a full-width at half maximum (FWHM) of $0.8$~eV.}%
\label{fig:spectra}%
\end{figure}

We observed strong laser-like emission in forward geometry direction, with similar angular divergence as the pump FEL pulse, at 65.18~nm, 68.14~nm, 68.8~nm and 108.9~nm in Xe (see Table~\ref{tab:exp_lines}). Only the $108.9$ nm emission line had previously been observed \cite{KapteynPRL86,Yamakoshi:96}. In Kr, we observed emission at $54.0$ nm and a weaker emission at the previously observed wavelength of 90~nm \cite{KapteynPRA88}. Examples of single 
shot and averaged Xe emission spectra are presented in Fig. \ref{fig:spectra}, showing two sharp lasing lines at $65.18$~nm and $68.14$~nm.

%%%%%%%%%%%%%%%%%%%%%%%%%%%%%%%%%%%%%%%%%%%%%%%%%%%%%%%%%%%%%%%%%%%%%%%%%%%%%%
\begin{table}[!htb]%[H] add [H] placement to break table across pages
 \caption{\label{tab:exp_lines}Observed superfluorescent emission line wavelengths $\lambda_L$, widths $\Delta\lambda$ (FWHM), photon energies $E_L$ and energy widths $\Delta E$ (FWHM, limited by the spectrometer resolution except for the 68.14 nm line). Numbers in brackets give the systematic uncertainty of the measurements.}
 \begin{ruledtabular}
 \begin{tabular}{ccccc}
 \toprule
 
Gas & $\lambda_L$[nm] & $\Delta\lambda$[pm] & $E_L$[eV] & $\Delta E$[meV]\\
Xe & $65.18(20)$ & $<1.6$ & $19.02(6)$ & $<$0.46 \\
Xe & $68.14(20)$ & 1.5 to 1.7 & $18.20(5)$ & 0.41 to 0.46 \\
Xe & $68.8(2)$ & $<1.6$ & $18.20(5)$ & $<$0.46 \\
Xe & $109.3(5)$ & $<2.4$ & $11.34(5)$ & $<$0.25 \\
Kr & $54.0(1.4)$ & $<1.7$ & $22.96(58)$ & $<$0.72 \\
Kr & $90.3(1.4)$ & - & $13.77(25)$ & - \\
 \end{tabular}
 \end{ruledtabular}
\end{table}
%%%%%%%%%%%%%%%%%%%%%%%%%%%%%%%%%%%%%%%%%%%%%
The level scheme for Xe is depicted in Fig. \ref{fig:Xe_scheme} \cite{Becker_PRA89, Kammerling_JPB92, Jauhiainen_JPB95, Saito_JPJ97, Luhmann_PRA98, Carrol_JESRP02, Lablanquie_JPB02, Viefhaus_JPB05, Penent_PRL05}. Following $4d$ ionization, the inner-shell excited states Xe$^+$ $4d_{3/2}^{-1}$ and $4d_{5/2}^{-1}$ rapidly decay by Auger process (lifetime $\sim 6$~fs \cite{JurvansuuPRA01}) to a manifold of several different valence-excited Xe$^{2+}$ states. 
This results in a population inversion between several pairs of these states. Additionally, these dicationic states can be created by a single-photon valence double ionization. This valence double ionization route is substantial, amounting for $\sim10\%$ of the total. Details on levels A, B and C as well as the respective partial Auger and valence double ionization rates were obtained by electron-electron coincidence measurements \cite{Khalal_PRA17} and are discussed in \cite{Sup_Mat}. The measured partial occupation rates are input parameters for our theory. The pumping scheme for Kr is similar, except that the photoionization occurs in the $n=3$ shell, creating Kr$^+$ $3d_{3/2}^{-1}$ and $3d_{5/2}^{-1}$ states with Auger lifetimes of~$\approx 7.5$~fs \cite{Minnhagen_AF69, Marr_ADNDT76, Hayaishi_JPB90, Bredice_JQSRT00, Tamenori_JPB04, Palaudoux_PRA10, Zeng_JPB13}. Superfluorescent emission then emanates from Kr$^{2+}$ ions with two holes in the $n=4$ shell.
%%%%%%%%%%%%%%%%%%%%%%%%%%%%
%%%%%%%%%%%%%%%%%%%%%%%%%%%%%%%%%%%%%%%%%

\begin{figure}[!htb]%
\includegraphics[width=8.6cm]{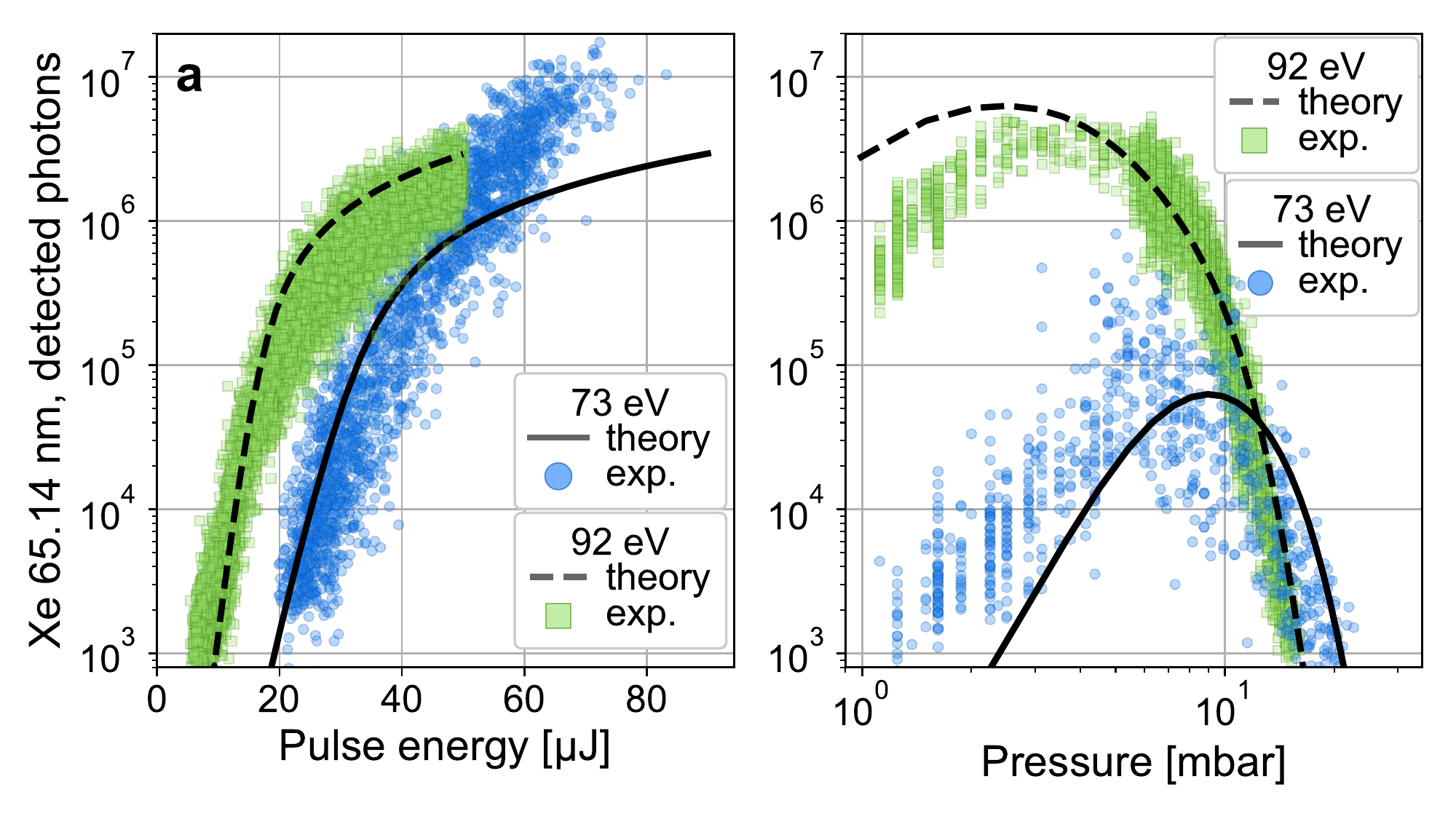}
\caption{(a): Symbols: measured intensity (detected photons) of the Xe 65.18 nm emission line as a function of FEL-pulse energy for 73 eV (at 7 mbar Xe) and 92 eV (at 3.5 mbar Xe) pump-photon energies. Lines: comparison with the model. (b): Symbols: pressure dependence of the same line for both pump-photon energies. The average pump-pulse energy was 30~$\mu$J for 73~eV and 75~$\mu$J for 92~eV. Lines: comparison with the model.}
\label{fig:gain_curves}%
\end{figure}
%%%%%%%%%%%%%%%%%%%%%%%%%%%%%%%%%%%%%%%%
We now focus on the Xe emission lines at wavelengths of 65.18 and 68.14 nm. In Fig. \ref{fig:gain_curves}-a, the integrated intensity of the 65.18~nm Xe line is shown as a function of the FEL-pulse energy for a photon energy of 73~eV (92~eV) at a pressure of 7 mbar (3.5 mbar). These values were chosen to optimize the emission-line intensity (see Fig. \ref{fig:gain_curves}-b). A clear exponential increase of the emission strength over 4 orders of magnitude results from varying the pulse energy from 10 to 60 $\mu$J. For 92~eV pump-photon energy, the emission strength saturates for pulse energies above 30~$\mu$J, while for 73 eV pump-photon energy, saturation sets in for pulse energies exceeding 50~$\mu$J. The signal above 50~$\mu$J pulse energy for 92-eV pump-photon energy was saturated on the detector and is not shown. The pulse-to-pulse variation of the integrated line emission intensity spreads over more than one order of magnitude for a given FEL-pulse energy, which was measured upstream the interaction volume and transport optics. The scatter is due to pulse energy measurement uncertainty as well as pointing instabilities of the FEL, leading to partial clipping of the focused beam by the gas-cell apertures. The intensity of the Xe 68.14~nm line showed a similar behaviour with pump-pulse energy, pump-photon energy and pressure \cite{Sup_Mat}.

The large difference of gain between 73-eV and 92-eV pump photon energy is not only due to the difference in $4d$ photoionization cross sections ($\sigma_{abs}$=5.2 Mb for 73 eV and $\sigma_{abs}$=25 Mb for 92 eV \cite{Shannon_JPB77, Becker_PRA89}), but also to differences in partial occupation rates of the upper and lower states of the emission lines (see Fig. \ref{fig:Xe_scheme}), as revealed by our electron-electron coincidence measurements in Xe. Single-photon, double-ionization of the valence orbitals plays a significant role in the occupation of the superfluorescence states: At 73 eV this pathway amounts to $\approx 15\%$ of the occupation of the upper state B, and $\approx 30\%$ of the lower states. At the peak of the $4d$ giant resonance, its relative contribution is smaller. Valence double ionization de facto reduces the population inversion created by the $4d$ ionization pathway, since this process tends to majorly populate energetically lower lying Xe$^{2+}$ states (see Fig. \ref{fig:Xe_scheme} and table S-1 in \cite{Sup_Mat}). This explains the larger gain of the 65 nm line at 92 eV and 100 eV compared to that at 73 eV. In addition, at 92~eV, a conjugate shake-up Auger decay, which is energetically inaccessible at 73 eV \cite{Hayaishi_PRA91}, slightly enhances the population of the upper state of the 65 and 68 nm lines.
%%%%%%%%%%%%%%%%%%%%%%

Insight to the amplification process can be gained through our theoretical model \cite{andrei}. The approach is based on a quantized treatment of both the emitted field and the atomic system and predicts the emitted intensity profile in the temporal and spectral domains for the ensemble average \cite{Sup_Mat}. Being fully quantized, it enables us to capture the cross-over from spontaneous emission through amplified spontaneous emission to superfluorescence. It is restricted here to a quasi one-dimensional pump geometry (single $k$-vector of the emission). Although we are facing a complex level structure of common emission upper states, we only consider a single transition (see Fig. \ref{fig:Xe_scheme}). This approximation thus does not take into account the competition between two transitions sharing the same upper state, but should be sufficient for capturing the emission dynamics before the system saturates and Rabi oscillations occur. Cascading superradiant processes, such as FEL induced multi-level superfluorescence in He \cite{doi:10.1063/1.89595, doi:10.7566/JPSJ.84.054301}, would need a more sophisticated level scheme. The pump process by the FEL and the Auger decay are modeled as incoherent processes according to the experimental cross sections and rates (Auger lifetime $1/\gamma_a = 6$~fs and partial occupation rates of Fig. \ref{fig:Xe_scheme}). Depending on the mean FEL intensity on target, the 80-fs pulses (modeled as Gaussian temporal pulses) prepare the atom in core-excited states on the time scale of 10-100 fs. The travel time of the pump pulse through the 4.5 mm long cell, $\sim 15$ ps, is short compared to the long radiative lifetime ($1/\Gamma_{sp}\sim 1$~to~$4$~ns) of the upper lasing state. Thus, a prompt population inversion is established, with $\sim10^{8}$ emitters in a medium of 20~$\mu$m cross sectional radius. The model accounts for a non-resonant absorption of the emitted field. Fitting the exponential decrease of intensity at high pressures, we estimated absorption cross-sections of the 65.18 and 68.14 nm emission lines to $\sigma_{absF} = 80$~Mb at 92 eV and $\sigma_{absF} = 60$~Mb at 73 eV pump photon energy. In addition, given the high FEL intensities, decoherence and depopulation by sequential multi-photoionization are included in the model for all involved levels, assuming cross sections for the neutral atom to be valid. At the considered pressures and electron densities, electron collisions were not assessed as critical \cite{KapteynPRL86}. The theory solves for the occupations of the Xe ground state, the $4d^{-1}$ inner-hole state, the lower and upper levels of the emission line as well as for the temporal and spectral properties of the outgoing radiation (see \cite{Sup_Mat}). %For the emitted electric field, a generalized correlation function $S(z_1,z_2,\tau)$ of the atomic system, measuring the spatial correlation of the coherences (polarization) at equal time point $\tau$ is used to calculate the temporal and spectral properties of the outgoing radiation.

Despite the level scheme simplification and the 1D approximation, the theory reproduces well the emission yield as a function of pump energy for both 73 and 92 eV photon energies (Fig. \ref{fig:gain_curves}-a). To compare with the experiment, we assumed a total detection efficiency of 5\%.
The pressure dependence of the yield is qualitatively reproduced by the theory (see Fig. \ref{fig:gain_curves}-b). %Our model assumes the ``cold" absorption cross section of neutral Xe for the emission lines transmitted through the cell. 
For a quantitative comparison, a 2D treatment of the superfluorescence as well as a kinetic code following the evolution of the transient plasma and its opacity would need to be employed.
%%%%%%%%%%%%%%%%%%%%%%%%%%%%%%%
\begin{figure}[!htb]%
\includegraphics[width=8.6cm]{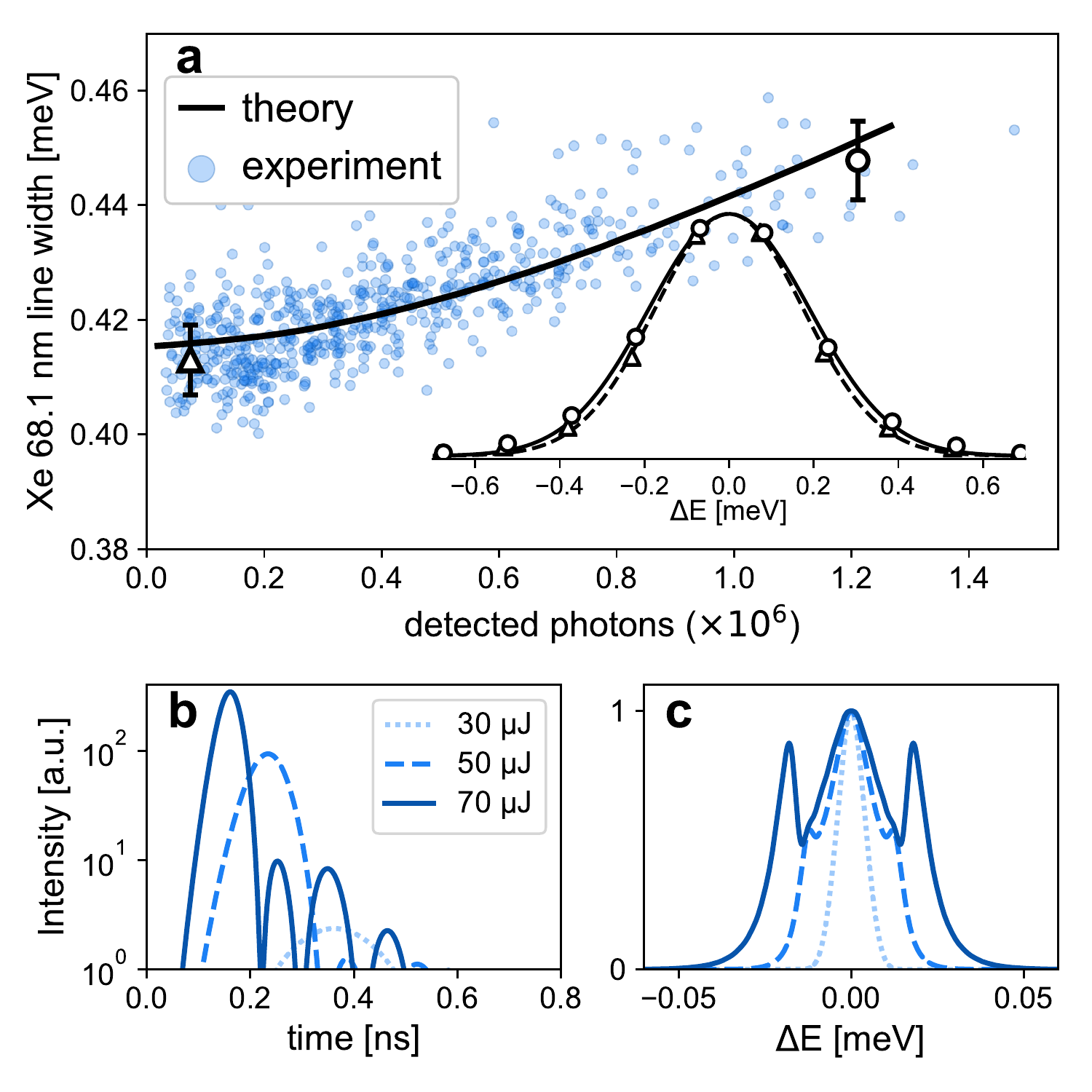}
\caption{(a) Measured line width of the Xe 68.14 nm emission line (FWHM of a Gaussian fit) as a function of emission yield. The signal is accumulated over three consecutive FEL pulses of 73 eV photon energy, each separated by 1~$\mu$s. During this scan, the pressure in the cell was varied between 30 and 44 mbar and showed no correlation with the line width. The error bars on the black triangle and circle show the standard deviation of the line width resulting from the fitting procedure. The corresponding normalized spectra (symbols) and fits (lines) to these data points are shown in the inset. Triangles: $\sigma=0.413\pm0.006$ meV and $7.4\times10^4$ detected photons. Circles: $\sigma = 0.448\pm 0.007$ meV and $1.2\times10^6$ detected photons. The black line is the theoretically determined line width taking into account the spectrometer resolution of 0.4 meV. (b) Theoretical time traces of the Xe 65.18 nm line for various pump-pulse energies, with the same parameters as in Fig. \ref{fig:gain_curves}-a and 73 eV photon energy. The FWHM peak width $\tau_W$ and delay time $\tau_D$ of the first emission peak are $\tau_W=$200 ps, $\tau_D=$360 ps for 30 $\mu$J, $\tau_W=$85 ps, $\tau_D=$230 ps for 50 $\mu$J and $\tau_W=$50 ps, $\tau_D=$160 ps for 70 $\mu$J. (c) Normalized theoretical spectral intensities.}
\label{fig:linewidth}%
\end{figure}
%%%%%%%%%%%%%%%%%%%%%%%%%%%%%%%

In the temporal domain, our theoretical approach reproduces well the typical features of superfluorescence: Fig. \ref{fig:linewidth}-b shows calculated time traces of the Xe 65.18~nm emission line obtained with the same parameters as in Fig. \ref{fig:gain_curves}-a for several pump-pulse energies at 73~eV pump-photon energy. Increasing the pump-pulse energy from 30 to 70~$\mu$J results in an effective increase of emitters from 3$\times 10^8$ to 6$\times 10^8$, and a decrease in the delay times $\tau_D$ from 360 to 160 ps, along with a decrease of the pulse duration $\tau_W$ from 200 down to 50 ps. Notably, the width $\tau_W$ and delay $\tau_D$ of the first emission peak as determined from our novel theoretical approach are a factor 3 to 5 larger as compared to the values of the phenomenological, analytical theory \cite{1976'MacGillivray}. For 70 $\mu$J (high within saturation), the typical ringing phenomenon is visible in the temporal intensity average. For nanosecond plasma-based pump sources, Kapteyn et al. \cite{KapteynPRA88} were able to measure time structures of the Xe 109 nm emission line for several output energies. The pulse duration of the emission was in the range of 600 to 1200 ps. At high emission intensity, a double-pulse structure of the emission was observed, with a peak separation of 800 ps. At the highest intensity the two peaks collapsed into a single emission peak of 450 ps duration. The evidence of this previous work points towards a cross-over from amplified spontaneous emission to superfluorescence.

In the spectral domain, we measured a quasi-linear broadening of the Xe 68.14 nm emission line as a function of its intensity (Fig. \ref{fig:linewidth}-a). This broadening is at the limit of the resolution of our setup (0.4 meV) and could only be observed in 3\textsuperscript{rd} diffraction order. Fig. \ref{fig:linewidth}-c shows the spectral output of our calculations. An increase of the number of emitters results in a broader emission line and the emergence of side bands. The spectral broadening corresponds to a decrease of the collective emission time (related to the width of the temporal peak $\tau_W$), and the side bands are manifestations of the coherent temporal ringing of the intensity in the spectral domain. The theoretically determined width, accounting for the spectrometer resolution, is shown as a black solid line in Fig. \ref{fig:linewidth}-a. While the side bands are not resolved in the experiment, the theoretical line width quantitatively matches the experimentally observed trend. Thus, the measured line broadening with emission yield is a strong indication of superfluorescence.\\

We presented experimental and theoretical data that underline the superfluorescent character of several Xe (Kr) lines prepared in a population-inverted state following $4d$ ($3d$) shell ionization by an FEL and subsequent Auger decay. Exponential growth of the emission yield as a function of pump-pulse energy was demonstrated over 4 orders of magnitude, reaching saturation with more than $10^7$ detected photons ($10^8$ to $10^9$ emitted photons). The line width of the emission showed an increase with the emission yield that was, within the experimental resolution, quantitatively predicted by our comprehensive theoretical approach. Collective emission times of the order of 100 ps are predicted and feature the typical ringing phenomenon, in line with early experiments on the temporal emission profiles of the Xe 109 nm line \cite{KapteynPRA88}. %The experimental findings therefore point to a strong evidence for XUV superfluorescence. 
Compared to other sources, such as FELs or high brilliance table-top XUV lasers approaching the carbon $1s$ edge  \cite{PhysRevA.89.053820,Rockwood:18} with ps \cite{PhysRevA.79.023810} and recently sub 100 fs \cite{sebban} duration, the demonstrated Auger-pumped superfluorescence source will not be competitive yet. Schemes could however be envisioned to shorten the pulse duration. Ionizing the upper lasing state with a short, time-delayed laser pulse would shrink the superfluorescence time, albeit at downscaled photon number. Coherent, optical quantum control schemes, such as recently suggested \cite{PhysRevA.88.053849,PhysRevA.94.023821}, could also be adapted to such a source. The fact that saturation could be reached  with optical pump sources \cite{Sher:87} is very appealing. The very good photon energy reproducibility even in the absence of a high-resolution setup can deliver lines for photonic studies with e.g. photoelectrons with an extremely narrow and well defined energy. The geometry of the gain medium could be optimized, allowing for a large number of emitters at higher solid angle \cite{PhysRevA.23.1334}. Auger- and Coster-Kronig pumped systems have been theoretically studied for only a few atoms \cite{Mendelsohn:85,0022-3700-16-24-023}, but a systematic search through other atomic or molecular gain media could lead to more XUV emission wavelengths.\\

%%%%%%%%%%%%
We acknowledge the technical and scientific teams at FLASH, in particular Dr. K. Tiedtke, Dr. A. Sorokin and Dr. R. Treusch for their support during the experiment. We acknowledge the Max Planck Society for funding the development and the initial operation of the CAMP end-station within the Max Planck Advanced Study Group at CFEL and for providing this equipment for CAMP@FLASH. The installation of CAMP@FLASH was partially funded by the BMBF grants 05K10KT2, 05K13KT2, 05K16KT3 and 05K10KTB from FSP-302. A.R. and D.R. are supported by the Chemical Sciences, Geosciences, and Biosciences Division, Office of Basic Energy Sciences, Office of Science, U.S. Department of Energy, Grant No. DE-FG02-86ER13491. B.E., C.B., and D.R. also acknowledge support through the Helmholtz Young Investigator program. Z.Y. acknowledges financial support by SFB 1073, project C02 from DFG. The coincidence experiments were performed at SOLEIL synchrotron; we are grateful to N. Jaouen and SEXTANTS team for help during the measurements, and to SOLEIL staff for stable operation of the storage ring. M. A. K. acknowledges the support of the Labex Plas@Par managed by the Agence Nationale de la Recherche, as part of the ``Programme d'Investissements d'Avenir" under Reference No. ANR-11-IDEX-0004-02.

%\bibliography{XUV_Superfluorescence_biblio}


\end{document}

%\preprint{}

\title{Supplementary Material to Extreme Ultraviolet Superfluorescence in Xenon and Krypton}


\author{L. Mercadier}
\email{laurent.mercadier@xfel.eu}
\affiliation{Max Planck Institute for the Structure and Dynamics of Matter, 22761 Hamburg, Germany}
\affiliation{European XFEL, 22869 Schenefeld, Germany}
\author{A. Benediktovitch}
\affiliation{Deutsches Elektronen-Synchrotron (DESY), 22761 Hamburg, Germany}
\author{C. Weninger}
\affiliation{Max Planck Institute for the Structure and Dynamics of Matter, 22761 Hamburg, Germany}
\author{M. A. Blessenohl}
\affiliation{Max-Planck-Institut f\"ur Kernphysik, 69117 Heidelberg, Germany}
\author{S. Bernitt}
\affiliation{Institut f\"ur Optik und Quantenelektronik, Friedrich-Schiller-Universit\"at Jena, 07743 Jena, Germany}
\affiliation{Max-Planck-Institut f\"ur Kernphysik, 69117 Heidelberg, Germany}
\author{H. Bekker}
\affiliation{Max-Planck-Institut f\"ur Kernphysik, 69117 Heidelberg, Germany}
\author{S. Dobrodey}
\affiliation{Max-Planck-Institut f\"ur Kernphysik, 69117 Heidelberg, Germany}
\author{A. S\'anchez-Gonz\'alez}
\affiliation{Department of Physics, Imperial College London, London SW7 2AZ, United Kingdom}
\author{B. Erk}
\affiliation{Deutsches Elektronen-Synchrotron (DESY), 22761 Hamburg, Germany}
\author{C. Bomme}
\affiliation{Deutsches Elektronen-Synchrotron (DESY), 22761 Hamburg, Germany}
\author{R. Boll}
\affiliation{Deutsches Elektronen-Synchrotron (DESY), 22761 Hamburg, Germany}
\author{Z. Yin}
\affiliation{Deutsches Elektronen-Synchrotron (DESY), 22761 Hamburg, Germany}
\affiliation{Max Planck f\"ur biophysikalische Chemie, 37077 Göttingen, Germany}
\author{V. P. Majety}
\affiliation{Max Planck Institute for the Structure and Dynamics of Matter, 22761 Hamburg, Germany}
\author{R. Steinbr\"ugge}
\affiliation{Max-Planck-Institut f\"ur Kernphysik, 69117 Heidelberg, Germany}
\author{M. A. Khalal}
\affiliation{Laboratoire de Chimie Physique - Mati\`ere et Rayonnement, Universit\'e Pierre et Marie Curie, F-75231 Paris Cedex 05, France}
\author{F. Penent}
\affiliation{Laboratoire de Chimie Physique - Mati\`ere et Rayonnement, Universit\'e Pierre et Marie Curie, F-75231 Paris Cedex 05, France}
\author{J. Palaudoux}
\affiliation{Laboratoire de Chimie Physique - Mati\`ere et Rayonnement, Universit\'e Pierre et Marie Curie, F-75231 Paris Cedex 05, France}
\author{P. Lablanquie}
\affiliation{Laboratoire de Chimie Physique - Mati\`ere et Rayonnement, Universit\'e Pierre et Marie Curie, F-75231 Paris Cedex 05, France}
\author{A. Rudenko}
\affiliation{J. R. Macdonald Laboratory, Department of Physics, Kansas State University, Manhattan, KS 66506, USA}
\author{D. Rolles}
\affiliation{Deutsches Elektronen-Synchrotron (DESY), 22761 Hamburg, Germany}
\affiliation{J. R. Macdonald Laboratory, Department of Physics, Kansas State University, Manhattan, KS 66506, USA}
\author{J. R. Crespo L\'opez-Urrutia}
\affiliation{Max-Planck-Institut f\"ur Kernphysik, 69117 Heidelberg, Germany}
\author{N. Rohringer}
\email{nina.rohringer@desy.de}
\affiliation{Max Planck Institute for the Structure and Dynamics of Matter, 22761 Hamburg, Germany}
\affiliation{Deutsches Elektronen-Synchrotron (DESY), 22761 Hamburg, Germany}
\affiliation{Department of Physics, Universit\"at Hamburg, 20355 Hamburg, Germany}

%\thanks{}

%\date{\today}

\maketitle


\section{Experimental setup}
The experiment was performed at the CAMP end station at FLASH. Pulses of $\sim80 - 100$ fs duration at 10-Hz repetition rate were focused to a nominal spot of $5\times7$~$\mu$m$^2$ into a pressurized gas cell of 4.5 mm length along the beam axis, so that the excited gas has a pencil-shaped geometry of large aspect ratio. The photon energy was tuned to 73~eV, 92~eV and 100~eV for Xe studies, and 100~eV for Kr. Assuming a beamline transmission of  50\%, the maximum available pulse energy on target was 90~$\mu$J, corresponding to an intensity of $4\times10^{15}$~W~cm$^{-2}$. 
The gas cell of adjustable pressure was used in previous experiments and consisted of three sections delimited by four Kapton windows. The central section of 4.5-mm length contained gas at 1-100 mbar pressure, while the two outer sections were used for differential vacuum pumping, ensuring a residual pressure in the main chamber below $10^{-5}$~mbar. 
The transmitted pulse and stimulated emission were impinging on a high resolution spectrometer consisting of a rotatable parabolic Pt-coated grating with 2400 grooves/mm and a radius of curvature of 3~m. The diffracted light was imaged onto a CCD camera (Andor, model Newton) at a fixed angle of 15\degree~with the incoming arm. The selection of the diffraction order and wavelength range of observation was achieved by rotating the grating. Given the large focal length of the spectrometer, the maximum solid angle detectable was $1.4\times10^{-4}$~sr. The size of the excited medium was estimated from the focusing geometry of the spectrometer to 20 - 30 $\mu$m radius.

The spectrometer calibration consists of determining parameters A and B of the following equation, for small grating angles: $k\lambda=A\theta - Bx$ where $\lambda$ is the wavelength, $k$ is the diffraction order, $\theta$ is the grating angle with respect to 0\textsuperscript{th} order and $x$ is the lateral position on the CCD in pixels. To do so, we performed absorption spectroscopy measurements of He auto-ionizing resonances (see Zubek et al., J. Phys. B: At., Mol. Opt. Phys. \textbf{22}, 3411 (1989) and Domke et al., Phys. Rev. A \textbf{53}, 1424 (1996)). With 80 mbar of He in the gas cell and a photon energy of 73 eV, we recorded spectra of self-amplified spontaneous emission (SASE) at different diffraction orders. For each order, averaging over a few hundred shots results in a smooth spectrum showing absorption dips at characteristic energies that we used for calibration. We found an angle dispersion of $A=(14.58 \pm 0.07)$~nm/deg, and a dispersion on the detector of $B=(1.70 \pm 0.02)$~pm/pixel, which corresponds to a resolution power of $\sim 2\times10^4$ and an observable wavelength window of $\approx 3.5$~nm in 1\textsuperscript{st} order of diffraction. In 3\textsuperscript{rd} order, this translates to a resolution power of $\sim 6\times10^4$.
%The spectrometer had a resolution power of $\lambda / \Delta \lambda \approx 2\times10^4$, with an observable window of 3.5~nm (1.0~eV) around 65~nm (19~eV) in the first order of diffraction.
The grating angle was manually adjustable and allowed the selection of the central wavelength with an accuracy of $\pm 1.4$~ nm. %Two gas monitor devices (GMD) located upstream and downstream the interaction point measured the pulse energy of the XFEL beam, thus allowing transmission measurements in the gaseous medium.
The detected number of photons was derived from the integrated CCD counts by taking into account the gain, and the wavelength-dependent quantum efficiency of the detector according to the constructor.

\begin{figure}[htb]
\includegraphics[width=0.8\linewidth, scale=1]{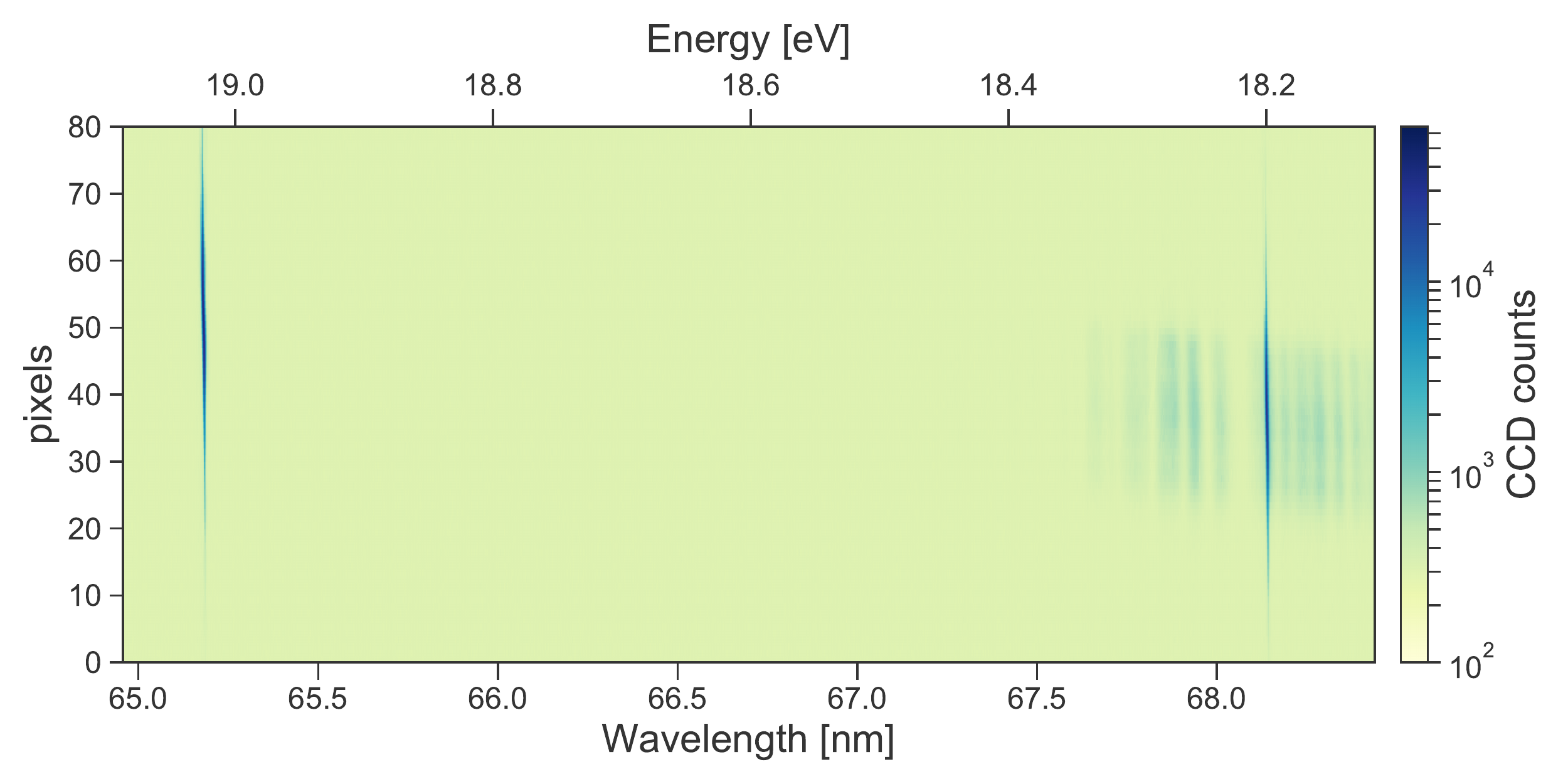}%
\caption{\label{fig:image} Spectral image recorded during a single FEL pulse exposure with 73 eV photon energy and 84 $\mu$J pulse energy. Xe pressure was 7 mbar. }
\end{figure}

Fig. \ref{fig:image} shows a typical spectral image obtained with a single FEL pulse of 73 eV photon energy and 84~$\mu$J pulse energy. The two Xe superfluorescence lines at 65.18 and 68.14 nm are observed. The spectrum obtained by vertical integration of the image corresponds to the blue line in Fig. 2 of the main text. The vertical axis is the spatial dimension while the horizontal axis is the energy-dispersed dimension. 

\section{Auger spectra and line identification}
%Detailed Auger-electron spectra with characterization of the correlation satellites at 73.7 eV photon energy were measured by Jauhiainen et al. \cite{Jauhiainen_JPB95} and Carroll et al. \cite{Carrol_JESRP02}, with determination of branching ratios between the different final channels. 73.7~eV is above the ionization thresholds of 67.548 and 69.537 eV \cite{King_JPB77} of the two $4d$ shell fine structure components $4d^{-1}_{5/2}$ and $4d^{-1}_{3/2}$, respectively. The Auger decay following $4d$ ionization leads to the formation of double (80\%) or triple (20\%) hole states in the $n=5$ shell \cite{Lablanquie_JPB02, Viefhaus_JPB05, Penent_PRL05, Kammerling_JPB92, Saito_JPJ97}. 

\subsection{Xenon}
The superfluorescent emission lines observed in Xe are assigned to transitions between Xe$^{2+}$ states which are efficiently populated by Auger decays. The gas - photon interaction at the used XFEL photon energies (73 and 92 eV) leads to the formation of Xe$^+$, Xe$^{2+}$ and Xe$^{3+}$. Both Xe$^{2+}$ and Xe$^{3+}$ levels can be populated by Auger decay, however analysis of coincidence spectra by Penent et al. [29] revealed that cascade emission of two Auger electrons following ionization in the $4d$ shell is the dominant process of the formation of Xe$^{3+}$ ions. These double decays only lead to the formation of ground state or low energy excitations of Xe$^{3+}$, with $5s_25p_3$ configurations. Radiative transitions between these low-energy excitations and the Xe$^{3+}$ ground state are too low in energy and superfluorescence in Xe$^{3+}$ ions can safely be excluded. The observed lasing transitions are therefore among Xe$^{2+}$ emission lines. The upper and lower states of the transitions could be identified from the literature and especially from the work by Juahiainen et al. [23]. Their assignment is presented in Table \ref{tab:lines}.
\begin{table*}[!htb]%[H] add [H] placement to break table across pages
\caption{\label{tab:lines}List of the observed XUV stimulated transitions in Xe, assigned from Ref. [23].}
 %\begin{ruledtabular}
\small 
\begin{tabular}{c@{\qquad}c>{\raggedright}p{3cm}c@{\qquad}clc}
\hline
\hline
$\lambda_L$~[nm] & \multicolumn{3}{c@{\qquad}} {Upper state} & \multicolumn{3}{c@{\qquad}} {Lower state}\\
& Name & Config. & Energy [eV] & Name & Config. & Energy [eV] \\
\hline
65.0479 & B &0.63 ($5s^2 5p^3 5d$) +~0.23~($5s^1 5p^5$) +~0.12~($5s^2 5p^3 6d$)
 & 20.2748 & $^3$P$_1$ & Xe$^{2+}$ $5p^{-2}$ ( $^3P_1$ ) & 1.2143 \\
\hline
68.2926 & B &0.63 ($5s^2 5p^3 5d$) +~0.23~($5s^1 5p^5$) +~0.12~($5s^2 5p^3 6d$)
& 20.2748 & $^1$D$_2$ & Xe$^{2+}$ $5p^{-2}$ ( $^1D_2$ ) & 2.1200\\
\hline
69.0400 & & $5s^2 5p^3(^2D\degree)5d$ $^3 P\degree_1$ & 19.1728  & $^3P_1$ & Xe$^{2+}$ $5p^{-2}$ ( $^3P_1$ ) & 1.2143 \\
\hline
108.8954 & C & $5s^0 5p^6$ ($^1$S$_0$) & 26.1430 & A & 0.65 ($5s^2 5p^3 5d$) +~0.25~($5s^1 5p^5$) & 14.7573\\
\hline
\hline
 \end{tabular}
 %\end{ruledtabular}
\end{table*}

\begin{figure}[!htb]
	\centering
		\includegraphics[width=8.6cm]{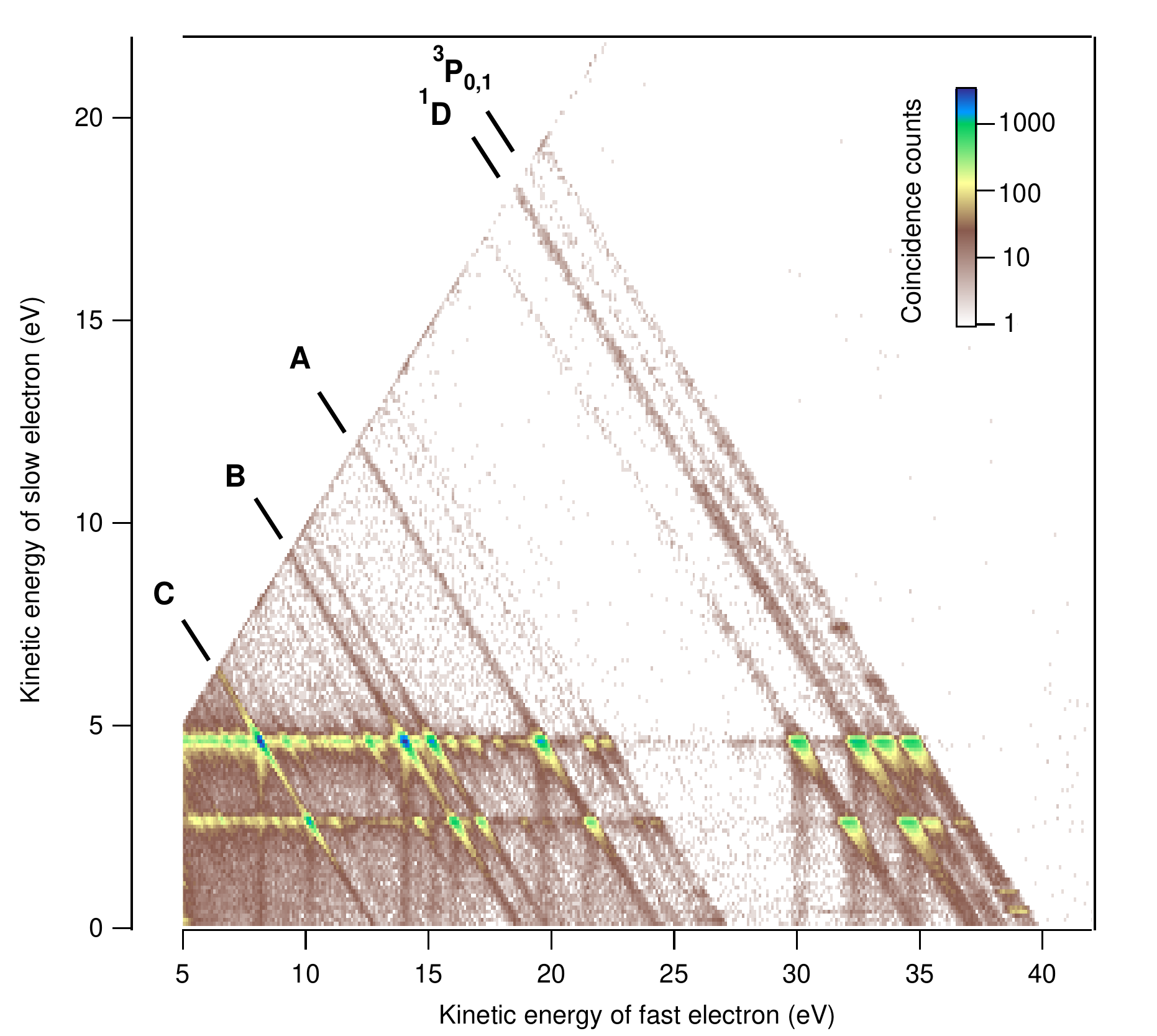}
	\caption{Energy correlation map between two electrons detected in coincidence upon photoionization of Xenon atoms with 73 eV photons. The map represents the kinetic energy of the slower electron of the pair (y axis) versus the faster one (x axis). Intensity is encoded in log scale. The two horizontal lines at 2.6 and 4.6 eV are due to the $4d$ photoelectrons. Diagonal lines diagonal lines correspond to formation of the different Xe$^{2+}$ final states. The Xe$^{2+}$ levels involved in the lasing lines are indicated with the same convention as in Table \ref{tab:lines}. See text for further details.}
	\label{fig:coincidences}
\end{figure}

\begin{figure}[!htb]
	\centering
		\includegraphics[width=8.6cm]{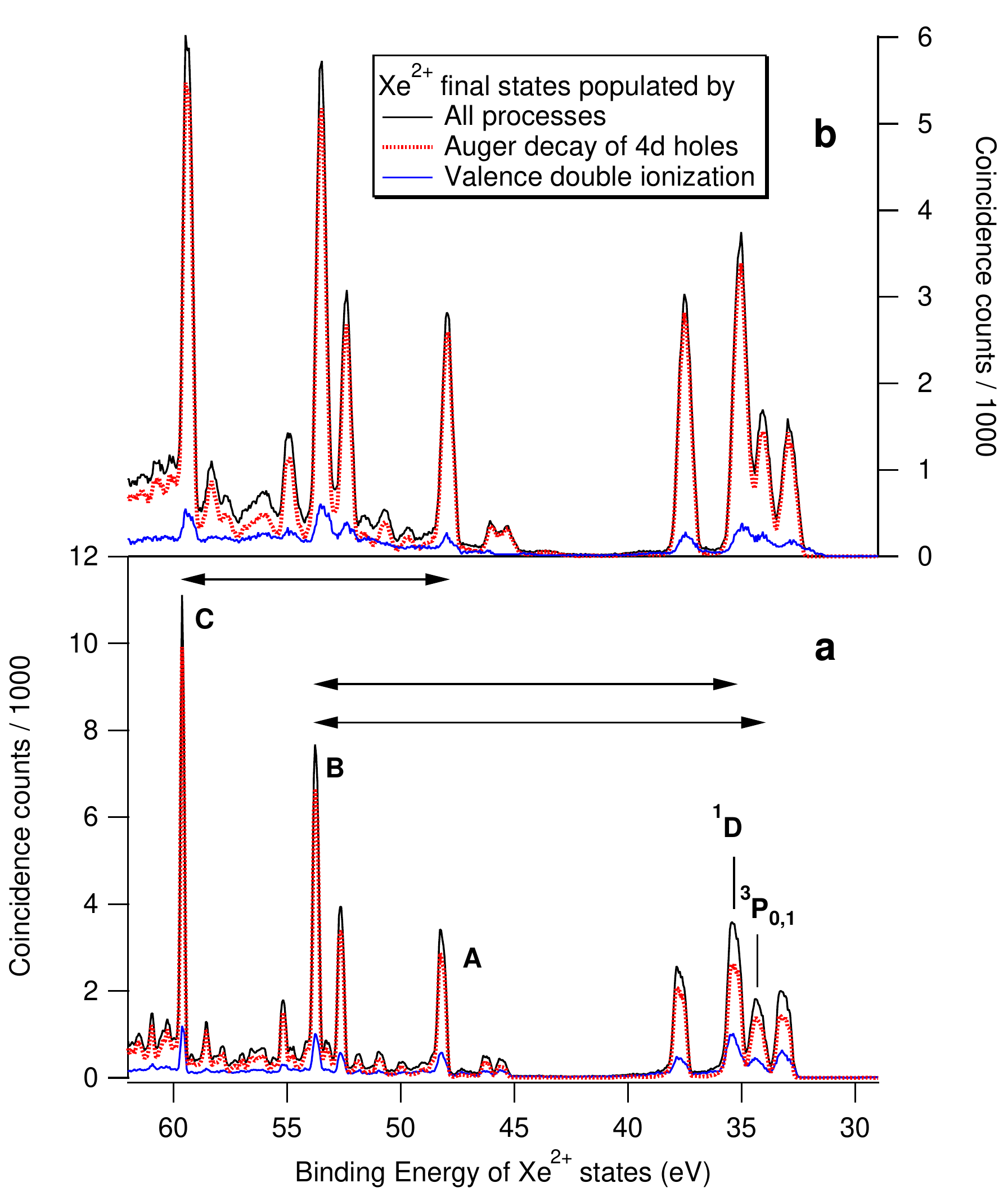}
	\caption{Population of Xe$^{2+}$ states by (a) 73 eV and (b) 92 eV incoming photon energies. The horizontal axis shows the sum energy of the two electrons involved in the formation of the Xe$^{2+}$ states. The upper and lower lasing states are shown as well as the different contributing processes ($4d$ hole + Auger decay or valence double ionization). Note that the broader lines observed at the higher photon energy reflect the decreased resolution when the electron kinetic energy increases.}
	\label{fig:auger_spectra}
\end{figure}

In order to get more details on the population of these Xe$^{2+}$ states, electron-electron coincidence measurements were performed with the HERMES magnetic bottle spectrometer [29, 31] at the SEXTANTS beamline of the synchrotron SOLEIL. Coincidence spectra were recorded for 73 eV and 92 eV photon energies for Xe, and 100 eV photon energy for Kr. Results for Xe are presented in Fig. \ref{fig:coincidences} and \ref{fig:auger_spectra}. 

First, the branching ratios for the population of  Xe$^+$, Xe$^{2+}$ and Xe$^{3+}$ species was estimated, by comparing the energy spectra of electrons detected in coincidence with 0, 1 or 2 other electrons. These figures take into account the estimated detection efficiency $f$ of the electrons which slowly decreases as a function of the kinetic energy of the photoelectron $E_c$,  from  $f = 71\%$ for $E_c$ = 4~eV to $f = 58\%$ for $E_c$ = 80~eV. The obtained Xe ion ratios are presented in Table \ref{tab:ratios}. They compare favorably with values from the literature which have been obtained by direct measurement of the ions yields as a function of the photon energy [20].

Second, the population of the Xe$^{2+}$ final states were investigated in more details, by looking at the energy correlation between the 2 emitted electrons. The energy correlation map between the two electrons emitted upon Xe$^{2+}$ formation by 73 eV photons is presented in Fig \ref{fig:coincidences}. The diagonal lines correspond to formation of the different Xe$^{2+}$ final states. The Xe$^{2+}$ levels involved in the lasing lines are indicated with the same convention as in Table \ref{tab:lines}. The two horizontal lines originate from the $4d$ photoelectrons which are observed here in coincidence with the corresponding Auger electrons. Intensity all along the diagonal lines outside of these $4d$ contributions come from double photoionization in the valence shell. This double photoionization can be a direct or a cascade one, as evidenced respectively by the weak intensity all along the diagonal lines and by the spots on these lines, and as was first shown by Eland et al. (Eland et al, Phys. Rev. Lett. \textbf{90}, (2003) 053003). Weak vertical structure originates from false coincidences between Auger electrons and slow energy electrons. From Fig. \ref{fig:coincidences}, we extract the relative contributions of the different processes to the population of the Xe$^{2+}$ final states and especially to the upper and lower states of the observed superfluorescent transitions; they are shown in Fig. \ref{fig:auger_spectra} and Table \ref{tab:rel_pop}. 

 73 eV photon energy lies below the $4d$ satellite threshold and it is only the decay of the $4d_{5/2}$ and $4d_{3/2}$ holes which contributes to the Auger decay at this photon energy. At 92 eV the $4d$ satellites to $4d$ main lines ratio is estimated to $11\pm1\%$. These satellites decay mainly (73\%) by a double Auger path and contribute only marginally to Xe$^{2+}$ population [25]. Population inversion between the superfluorescent states is thus only weakly affected by the contribution of the $4d$ satellite states, which are found to populate relatively more Xe$^{2+}$ excited states than the $4d$ main lines. The main difference between the two photon energies originates from the contribution of the valence double ionization process, which is observed to populate mainly low energy Xe$^{2+}$ final states. Valence double photoionization contributes less at 92 eV photon energy which is close to the maximum of the Xe $4d$ giant resonance [20]. As a conclusion, our coincidence studies show that population inversion is due to the Auger decay of the $4d$ inner-shell holes, and that it is reduced at 73 eV photon energy because of the higher contribution of the valence double ionization path at this photon energy. 

\begin{table*}[!htb]%[H] add [H] placement to break table across pages
\caption{\label{tab:ratios}Branching ratios in \% for the formation of Xe$^{n+}$ ions at the two photon energies examined in our coincidence experiments. They are obtained from the spectra of electrons detected in coincidence with 0, 1 or 2 other electrons. Estimated error bars are $\pm1$ \%. They are compared with values estimated from Ref. [20].}

 %\begin{ruledtabular}
\small 
 \begin{tabular}{c@{\qquad}cc@{\qquad}cc}
 \hline
 \hline
% & \multicolumn{4}{c} {Branching ratios (\%)} \\
 & 73 eV & 73 eV \footnotemark[1] & 92 eV & 92 eV \footnotemark[1] \\
 \hline
 Xe$^+$ & 27 & 22 & 6 & 7 \\
 Xe$^{2+}$ & 59 & 62 & 65 & 67 \\
 Xe$^{3+}$ & 14 & 16 & 29 & 26 \\
 \hline
 \hline
 
 \footnotetext[1]{Ref. [20]}

 \end{tabular}
 %\end{ruledtabular}
\end{table*}

%\squeezetable
\begin{table}[!htb]%[H] add [H] placement to break table across pages
\caption{\label{tab:rel_pop}Relative contribution of $4d$ hole Auger decay (including $4d_{5/2}$, $4d_{3/2}$ and $4d$ satellites contributions) versus other processes (direct or cascade valence double ionization) for the population of the upper and lower states of the observed superfluorescent transitions. The last row shows the branching ratios, i.e. the proportion of the Xe$^{2+}$ upper and lower states relative to all Xe$^+$, Xe$^{2+}$ and Xe$^{3+}$ created by the photon - gas interaction. Note that the $^3P_1$ and the $^3P_0$ are not resolved in our experiment (see Fig. \ref{fig:coincidences}), and that the sum of the two is shown below ($^3P_{0,1}$).}

%\begin{ruledtabular}
\small 
\begin{tabular}{c@{\qquad}cc@{\qquad}cc@{\qquad}cc@{\qquad}cc@{\qquad}cc}
 \toprule
 \hline
\hline
& \multicolumn{2}{c@{\qquad}}{C} & \multicolumn{2}{c@{\qquad}} {B} & \multicolumn{2}{c@{\qquad}} {A} & \multicolumn{2}{c@{\qquad}} {$^1D_2$} & \multicolumn{2}{c@{\qquad}} {$^3P_{0,1}$}\\
 & 73 eV & 92 eV & 73 eV & 92 eV & 73 eV & 92 eV & 73 eV & 92 eV & 73 eV & 92 eV\\
 \hline
$4d$ holes & 87\% & 89\% & 85\% & 85\% & 82\% & 89\% & 71\% & 89\% & 74\% & 84\%\\
Valence double ionization & 13\% & 11\% & 15\% & 15\% & 18\% & 11\% & 29\% & 11\% & 26\% & 16\%\\
%\hline
%Branching ratios & 4.1\% & 4.7\% & 4.4\% & 4.7\% & 2.5\% & 2.4\% & 4.0\% & 4.0\% & 2.1\% & 1.6\%
Branching ratios & 5.8\% & 7.9\% & 6.2\% & 7.8\% & 3.5\% & 4.0\% & 5.7\% & 6.7\% & 3.0\% & 2.7\%\\
\hline
\hline
 \end{tabular}
 %\end{ruledtabular}
\end{table}

\subsection{Krypton}
In Kr, the same reasoning applies. The small line width shows that the upper state has a lifetime of $\approx400$~fs, thus superfluorescence occurs after the pulse has propagated through the gaseous medium. The 100-eV photon energy XFEL beam photoionizes the $3d$ shell with a cross-section of 1.5 Mb [33]. As in the case of Xe at 73 eV, 100 eV is below the shake-up limit. The core-excited states $3d^{-1}_{3/2}$ and $3d^{-1}_{5/2}$, with spin-orbit splitting of 1.26~eV, decay predominantly via Auger processes, forming double ($\approx 70\%$) or triple ($\approx 30\%$) hole states in the $n=4$ shell [33 - 37]. 

The simpler atomic structure of Kr with respect to that of Xe allows for ab-initio predictions of the states that correspond to the observed superfluorescent emission.
A recent study by Zeng et al. [39] using Flexible Atomic Code (FAC) (M. F. Gu, Can. J. Phys. \textbf{86}, 675 (2008)) shows that population inversion occurs in Kr$^{2+}$ but not in Kr$^{3+}$. Hence, the final excited states of Kr$^{2+}$ are potential candidates for the upper state of the transition observed. The paper of Zeng et al. gives the relative populations of different states of Kr$^{2+}$ and Kr$^{3+}$ after single Auger decay, cascade or double direct Auger decays, but does not give information on further photoemission decay. Therefore, we calculated transition rates and dipole moments in Kr$^{2+}$ using FAC. Population inversion is not sufficient to induce superfluorescence: the transition also requires a strong dipole moment and high transition rate. To take into account the strong electron correlation and configuration mixing, and remain consistent with the Zeng et al. approach, we considered a total of 34 configurations of Kr$^{2+}$: $4s^2 4p^4$, $4s^14p^5$, $4s^04p^6$, $4s^24p^25s^2$, $4s^24p^3nl$, $4s^24p^24d^1nl$, $4s^14p^4nl$, $4s^14p^34d^1nl$, $4s^14p^24d^2nl$ and $4s^04p^5nl$ ($nl=4d, 4f, 5s, 5p$ and $5d$). We could match two lines with significant dipole moments, for which the upper and lower state energies given by Zeng et al. show a sizable population inversion. Their characteristics are presented in Table \ref{tab:linesKr}. Although FAC accuracy for transition energy is limited to 1 to 2 eV, the calculated line energies are close to the transition observed. The transition from $4s^2 4p^3 4d^1$ to $4s^2 4p^4$ is the one observed by Minnhagen et al. [33] (last row of Table \ref{tab:linesKr}). We could not find any experimental match for the transition from $4s^1 4p^5$ to $4s^2 4p^4$ in the literature.

\begin{table*}[htb]% add [H] placement to break table across pages
\caption{\label{tab:linesKr}List of the best candidates for the observed XUV stimulated transition in Kr. The wavelength
$\lambda_L$, energy $E_L$, transition rate $\Gamma$ and dipole moment $\mu$ as well as the configurations, J number and energies 
of the levels were calculated using FAC. The Auger decay branching ratios BR as well as the difference $\delta$ between Auger 
decay to upper and lower states according to Ref. [39] are also indicated. The last row shows experimental fluorescence results by Minnhagen et al. [33] of the most probable transition we observed.}

 \begin{ruledtabular}
 \begin{tabular}{cccclccclcccc}
 \toprule
 $\lambda_L$~[nm] & $E_L$ [eV]& $\Gamma$~[$10^{10}$~s$^{-1}$] & $\mu$~[a.u.] &\multicolumn{4}{c} {Upper state} & \multicolumn{4}{c} {Lower state} & $\delta$~[$10^{13}$~s$^{-1}$]\footnotemark[1]\\
& & & & Config. & J & E [eV] & BR\footnotemark[1] & Config. & J & E [eV] & BR\footnotemark[1] \\
56.331 & 22.01 & 2.39 & 1.45 & $4s^2 4p^3_{3/2} 4d_{5/2}$ & 1 & 26.27 & 6.1\% & $4s^2 4p^4_{3/2}$ & 0 & 4.26 & 2.8\% & 1.28\\
54.135 & 22.90 & 2.31 & 1.35 & $4s_{1/2} 4p^2_{1/2}4p^3_{3/2}$ & 1 & 24.97 & 11.3\% & $4s^2 4p_{1/2} 4p^3_{3/2}$ & 2 & 2.07 & 7.0\% & 1.66\\
\hline
\multicolumn{2}{l} {From Ref. [34]:}\\
54.6687 & 22.6792 & & &$4s^2 4p^3(^2P^o)4d$~$^3D^o$ &1 & 23.337946 & & $4s^2 4p^4$~$^3P$ & 0 & 0.65872
\footnotetext[1]{Ref. [39]}

 \end{tabular}
 \end{ruledtabular}
\end{table*}

\section{Additional experimental data}

\begin{figure}[htb]
\includegraphics[width=1\linewidth, scale=1]{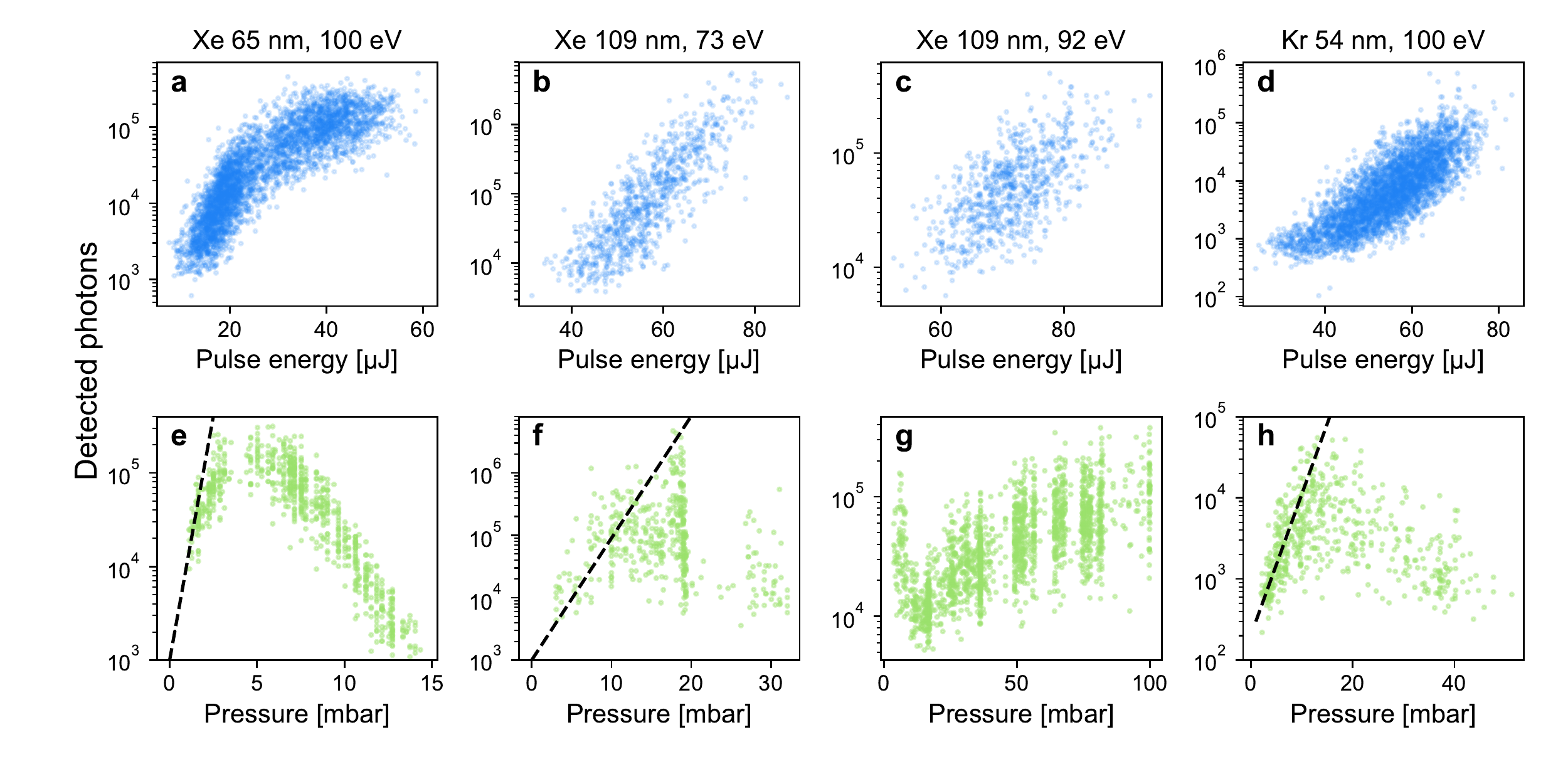}%
\caption{\label{fig:scans} Top row: Pump energy scans of observed superfluorescent lines. Average pressure is (a): 5.5 mbar, (b): 10 mbar, (c): 91 mbar, (d): 9.8 mbar. Bottom row: Pressure scans of the same lines. Average pump energy is (e): 50 $\mu$J, (f): 58 $\mu$J, (g): 80 $\mu$J, (h): 75 $\mu$J. The Kr line is observed in 2nd diffraction order, hence the lower detected number of photons with respect to the energy scan (d).}
\end{figure}

We present here pump energy and gas pressure scans of several emitted lines. The small-signal gain coefficient of the amplification process can be experimentally determined by the pressure dependence of the emission lines at fixed target length (bottom line of Fig. \ref{fig:scans} and Fig. 3-b in the main text). Fitting the exponential rise of the integrated line intensity (dashed lines in Fig. \ref{fig:scans}), we determined gain coefficients of the excited medium and summarized them in Table \ref{tab:gain_coeffs}. We note that the gain coefficient for the Xe 109 nm line pumped with 73 eV is $g'=0.8$~cm$^{-1}$~mbar$^{-1}$, which is similar to previous results (Sher et al. [50] observed $g'=0.73$~cm$^{-1}$~mbar$^{-1}$ for the same Xe 109 nm line). The behaviour of the Xe 109 nm line with 92 eV pump energy is somewhat different than the other lines, and no gain coefficient could be extracted. %The gain of the Xe 65.18 nm line is therefore more than three times larger than that of the Xe 109 nm line.The rise of the Xe 65.18 nm line intensity at 92 and 100~eV was too sharp and lasing was already significant at the smallest measured pressure of 1 mbar, preventing a measurement of the gain at those photon energies. We however estimate that the gain at 92 eV (100 eV) is at least three (two) times larger than that at 73 eV.

\begin{table}[!htb]%[H] add [H] placement to break table across pages
\caption{\label{tab:gain_coeffs}Gain coefficients $\alpha$ extracted from the pressure scans of various superfluorescent lines, as a function of FEL photon energy $E_{ph}$ and pulse energy $E_{FEL}$.}
 %\begin{ruledtabular}
 \begin{tabular}{cccc}
\hline
\hline
Line & $E_{ph}$ (eV) & $E_{FEL}$ ($\mu$J) & $\alpha$ (cm$^{-1}$~mbar$^{-1}$)\\
\hline
Xe 65.18 nm & 73 & 30 & 2.0\\
Xe 65.18 nm & 92 & 75 & 5.5\\
Xe 65.18 nm & 100 & 50 & 5.3\\
Xe 109 nm & 73 & 58 & 0.8\\
Xe 109 nm & 73 & 58 & 0.08\\
Kr 54 nm & 100 & 75 & 1.0\\
 \hline
 \hline
 \end{tabular}
 %\end{ruledtabular}
\end{table}

\begin{figure}[htb]
\includegraphics[width=0.5\linewidth, scale=1]{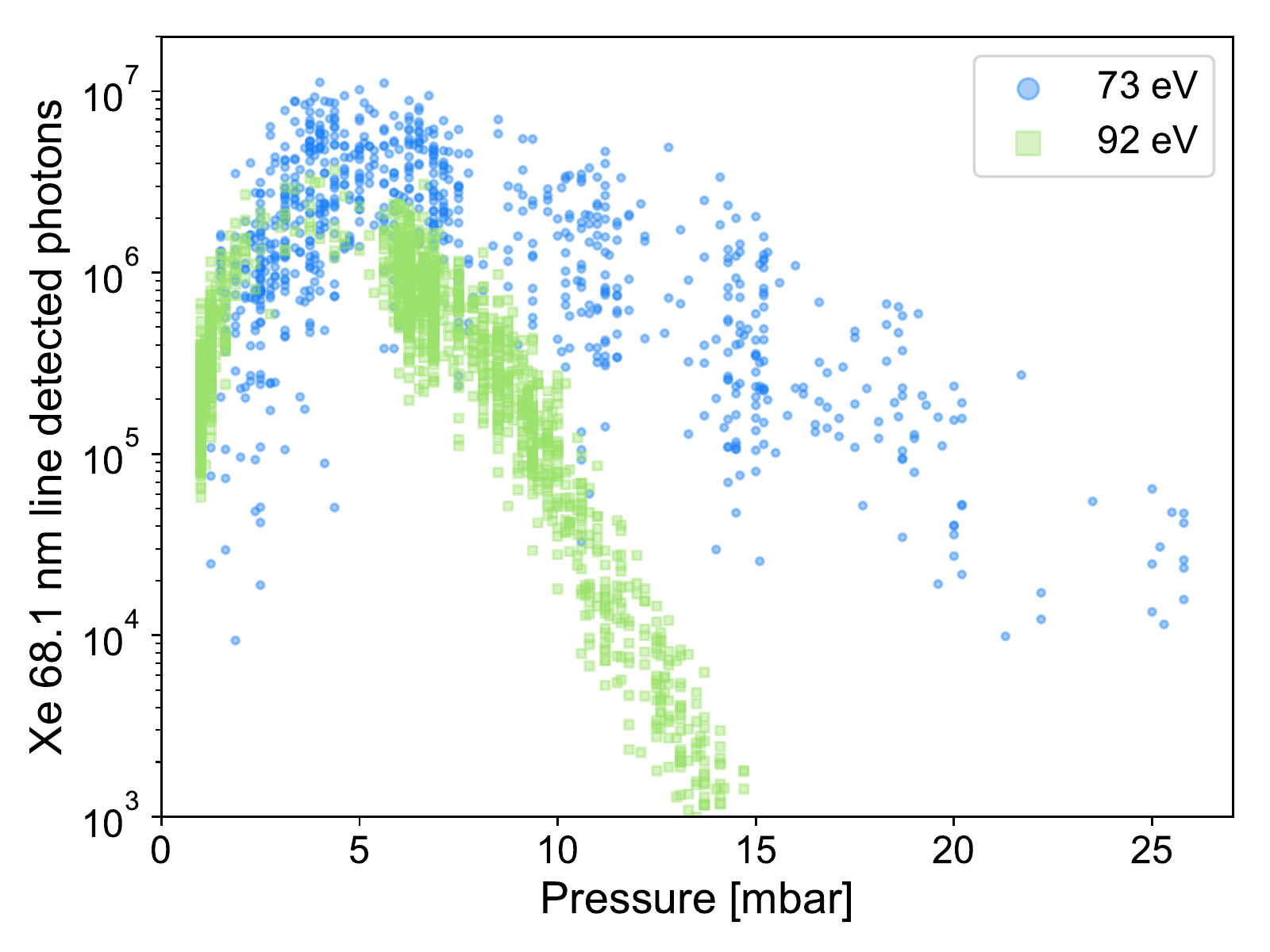}%
\caption{\label{fig:Xe68nm} Detected photons of the Xe 68.14 nm line as a function of pressure for 73 and 92 eV pump photon energies. Average pulse energy is 58 $\mu$J for 73 eV and 65 $\mu$J for 92 eV.}
\end{figure}

\clearpage
\section{Description of the superfluorescence model}

\subsection{Theoretical framework}

Here, we describe the few level model for Xe and provide the explicit forms for the time evolution of the populations, correlation function for atomic coherences and the two-point time correlation function for the field operators that gives the temporal and spectral properties of the emitted field. The procedure to obtain these equations is summarized, while a detailed derivation will be presented elsewhere.

Starting from the Hamiltonian (within the rotating wave approximation) for the quantized fields interacting with a set of uniformly distributed two-level atoms, the Heisenberg equations of motion for the field and atomic operators were derived. The field elimination was then performed to express the evolution of atomic operators for a given atom in terms of atomic operators of given and other atoms at previous time moments and field operators at initial time moment. Following this, Weisskopf-Wigner approximation was performed on the terms corresponding to the radiation reaction (spontaneous emission). In order to obtain numerical solutions with moderate computational effort and to allow for parameter studies, a one-dimensional approximation is performed. In addition, using the swept-pumping property leads to time evolution equations in the moving frame of the field. Finally, the time evolution equations for the field and atomic operators were used to evaluate various time and space correlation functions required to obtain the experimentally relevant observables. The higher order correlation functions appearing in the time evolution, beyond second order, were suitably factorized to obtain closed form equations.

In a real system, multiple atomic levels beyond the two lasing levels are involved. The feeding and the depletion of the lasing levels from/to other atomic levels due to pumping and non-radiative decay mechanisms was treated using suitable Lindbladian terms of a corresponding master equation. The influence of these terms on the evolution of the correlation functions was computed following Scully et al., \textit{Quantum Optics}, Cambridge (1997), Ch.8.

The sample in the experiment is assumed to be a quasi one-dimensional medium with a characteristic radius $R$. For a given three-dimensional density $n_3$, the one-dimensional density is given as $n_1 = n_3 \pi R^2$. The one-dimensional approximation introduces a solid angle parameter, denoted as $\Delta o$, the angle into which the atoms are emitting coherently. This approximation can be justified for systems that have Fresnel number about 1, see [1] for further discussions. To take this into account we divide the medium into subregions (along the transverse direction) with Fresnel number 1, take $\Delta o$ according to dimensions of the subregion and assume radiation from subregions to be independent.

% \section{Few level model of Xenon}
\begin{figure}[htbp]
\centerline{
  % Resize it to 5cm wide.
  \resizebox{10cm}{!}{
    \begin{tikzpicture}[
      scale=0.5,
      level/.style={thick},
      virtual/.style={thick,densely dashed},
      trans/.style={thick,->,shorten >=2pt,shorten <=2pt,>=stealth},
      redtrans/.style={thick,->,shorten >=2pt,shorten <=2pt,>=stealth,color=red},
      classical/.style={thin,double,<->,shorten >=4pt,shorten <=4pt,>=stealth}
    ]
    % Draw the energy levels.
    \draw[level] (0cm,0em) -- (-2cm,0em) node[left] {\ket{0}};%-5 for D level picture
    \draw[level] (0cm,13em) -- (2cm,13em) node[midway,above] {\ket{c}};
    \draw[level] (6cm,8em) -- (8cm,8em) node[midway,above] {\ket{e}};
    \draw[level] (6cm,5em) -- (8cm,5em) node[midway,below] {\ket{g}};
%    \draw[virtual] (6cm,-3em) -- (8cm,-3em) node[midway,below] {\ket{D}};
%     % Draw the transitions.
    \draw[trans] (-1cm,0em) -- (1cm,13em) node[midway,left] {$\gamma^0_{\text{pump}}$};
    \draw[trans] (1cm,13em) -- (6cm,8em) node[midway,right] {$\quad\gamma_{A} b_e$};
    \draw[trans] (1cm,13em) -- (6cm,5em) node[below,left] {$\gamma_{A} b_g \quad$};
%    \draw[trans] (1cm,13em) -- (6cm,-3em) node[below,left] {$\gamma_{A} b_D \quad$};
    \draw[redtrans] (7cm,8em) -- (7cm,5em) node[midway,right] {\Om{F}};    
    \draw[trans] (1.8cm,13em) -- (4cm,20em) node[above,right] {$\gamma^c_{\text{pump}}$};
    \draw[trans] (7.8cm,8em) -- (10cm,15em) node[above,right] {$\gamma^e_{\text{pump}}$};
    \draw[trans] (7.8cm,5em) -- (10cm,12em) node[above,right] {$\gamma^g_{\text{pump}}$};
%    \draw[trans] (7.8cm,-3em) -- (10cm,4em) node[above,right] {$\gamma^D_{\text{pump}}$};
    \end{tikzpicture}
  }
}
\caption{Few level model used to simulate Xe superfluorescence. Description is in the text below.}
\end{figure}
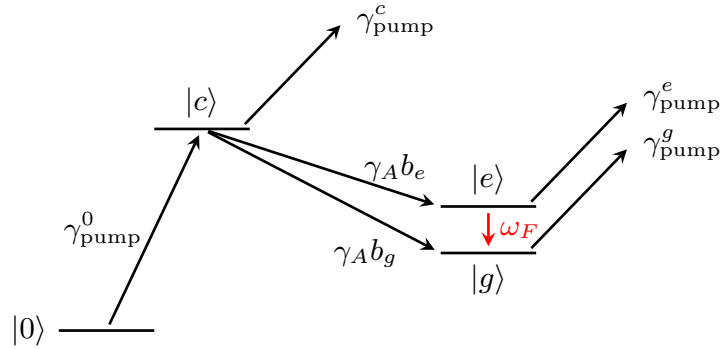

The emission at a given superfluorescence frequency, $\omega_{\text{F}}$, is modeled to be a consequence of the following process:
The ground state ($|0\rangle$) of Xe is pumped by a FEL pulse at central frequency, $\omega_{\text{pump}}$. This creates $4d^{-1}$ inner valence-hole state ($|c\rangle$) which decays by an Auger process into states that predominantly have $5s^{-1}$ and $5p^{-1}$ hole nature. These states constitute the superfluorescence levels $|g\rangle, |e\rangle$. The fluorescence rate between these levels is denoted by $\Gamma_{sp}$.
Further details on the superfluorescence levels leading to each of the emitted frequencies observed in the experiment can be found in the main text and in Table \ref{tab:lines}. 

The Auger lifetime ($1/\gamma_a$) of $|c\rangle$ is 6 fs. The branching ratios ($b_g,b_e$) into the individual lasing levels are estimated from coincidence measurements. The pumping frequency of 73 eV is in the vicinity of the so-called giant resonance feature. The photoionization cross sections of all the states involved are comparable and denoted as $\sigma_{abs}$.
% To model the overall depletion of the pump we additionally include a dummy level ($|D\rangle$) with a branching ratio $b_D = 1-b_g-b_e$. 

The pumping rate for a given state $|a\rangle$ is given by $\gamma_{pump}^a = \sigma_{abs} J$, where $J:=J_0(z,\tau) \mathcal{G}(z)$ is the pump flux, a function of 1-D space ($z$) and the retarded time ($\tau$) variables; $J_0(z,\tau)$ is the pump flux at beam waist located at the center of the medium and $\mathcal{G}(z) = \left( 1+\left( \frac{z}{z_R}\right)^2 \right)^{-1}$ is a correction factor to account for the Gaussian beam shape with $z_R$ being the Rayleigh range.

The evolution of pump flux and the populations of the ground state ($\rho_0$) and $4d^{-1}$ hole state ($\rho_c$) are given as:

\begin{equation}
\begin{aligned}
 \frac{\partial \rho_0(z,\tau)}{\partial \tau} &= -\sigma_{abs} \rho_0(z,\tau) J(z,\tau), \\
 \frac{\partial \rho_c(z,\tau)}{\partial \tau} &= -\gamma_A \rho_c(z,\tau) + \sigma_{abs} \left( \rho_0(z,\tau) -\rho_c(z,\tau) \right) \;J(z,\tau), \\ 
% \frac{\partial \rho_D(z,\tau)}{\partial \tau} &=  \gamma_A (1-b_e-b_g) \rho_D(z,\tau) - \sigma_{abs} \rho_D(z,\tau) J(z,\tau)\\
 \frac{\partial J_0(z,\tau)}{\partial z} &=  -\sigma_{abs} n_3 J_0(z,\tau). %(\rho_0 + \rho_c + \rho_{ee} + \rho_{gg} + \rho_D)(z,\tau)
\end{aligned}
\end{equation}

From energetic considerations, the emitted field can also be absorbed by the ground state with a cross-section $\sigma_{absF}$. In our current problem, we are in a regime where $\frac{1}{\Gamma_{sp}} \gg \frac{1}{\Gamma_{col}} \gg \tau_{pump}$ and the number of emitted photons is orders of magnitude lesser than the pump photons. Hence, we neglect the influence of the emitted field on the ground state and define an absorption kernel for the field solely based on the ground state populations at the end of pump pulse as
\begin{equation}
 \mathcal{A}(z,z') = e^{-\int\limits_{z'}^{z} dz'' \frac{\kappa(z'')}{2} },
\end{equation}
with 
\begin{equation}
 \kappa(z) = \sigma_{absF} \; \rho_0(z,\tau_{pump}).
\end{equation}

The time evolution for the populations of upper ($\rho_{ee}$), lower lasing levels ($\rho_{gg}$) and the two-point space correlation function of atomic coherences ($S(z_1,z_2,\tau)$) is given as:
\begin{equation}
\begin{aligned}
 \frac{\partial \rho_{ee} (z,\tau)}{\partial\tau} &= - \Gamma_{sp} \rho_{ee} - J \sigma_{abs} \rho_{ee} + \gamma_{A} b_{e} \rho_c - \frac{3 \Delta o}{8 \pi} \Gamma_{sp} n_1 \int\limits_{0}^z dz' 
 \mathcal{A}(z,z') S(z,z',\tau), \\
 \frac{\partial \rho_{gg} (z,\tau)}{\partial t} &=  \Gamma_{sp} \rho_{ee} - J \sigma_{abs} \rho_{gg} + \gamma_{A} b_{g} \rho_c  + \frac{3 \Delta o}{8 \pi} \Gamma_{sp} n_1 \int\limits_{0}^z dz' \mathcal{A}(z,z') S(z,z',\tau), \\
 \sigma(z,\tau) &= \frac{1}{2} \left( \rho_{ee} - \rho_{gg} \right), \\
 \frac{\partial S(z_1,z_2,\tau)}{\partial \tau} &= -\Gamma_{sp} S(z_1,z_2,\tau)\\
 &+ \underbrace{\frac{3 \Delta o}{8 \pi} \Gamma_{sp} n_1 \left( \sigma(z_1,\tau) \int\limits_0^{z_1} dz' \mathcal{A}(z_1,z') S(z',z_2,\tau) + \sigma(z_2,\tau) \int\limits_0^{z_2} dz' \mathcal{A}(z_2,z') S(z_1,z',\tau) \right)}_{\text{Stimulated emission}} \\
 &+ \underbrace{\frac{3 \Delta o}{8 \pi} \Gamma_{sp} \left( \sigma(z_1,\tau) \rho_{ee}(z_2,\tau) \Theta(z_1-z_2) \mathcal{A}(z_1,z_2) + \sigma(z_2,\tau) \rho_{ee}(z_1,\tau) \Theta(z_2-z_1) \mathcal{A}(z_2,z_1) \right)}_{\text{Spontaneous emission}}.
 \end{aligned}
\end{equation}

These objects can be used to compute the temporal properties of the emitted field and hence the gain dynamics. The spontaneous intensity and the stimulated intensity are computed as:
\begin{equation}
 \mathcal{I}(z,\tau) = \mathcal{I}_{sp}(z,\tau) + \mathcal{I}_{stld}(z,\tau)
\end{equation}
where

\begin{equation}
 \begin{aligned}
  \mathcal{I}_{sp}(z,\tau)   &= \frac{3 \Delta o}{32 \pi \lambda^2} \Gamma_{sp} n_1 \int\limits_{0}^{z} dz' \mathcal{A}(z,z') \rho_{ee}(z',\tau),\\
  \mathcal{I}_{stld}(z,\tau) &= \frac{3 \Delta o}{32 \pi \lambda^2} \Gamma_{sp} n_1^2 \int\limits_0^z dz_1' \mathcal{A}(z,z_1') \int\limits_0^z dz_2' \mathcal{A}(z,z_2') S(z_1',z_2',\tau),   
 \end{aligned}
\end{equation}

which lead us to the number of emitted photons:
\begin{equation}
 \mathcal{N}_{ph}(z) = \Delta o \;\pi R^2 \int\limits_0^\infty d\tau' \mathcal{I}(z,\tau'). 
\end{equation}

Finally, to study the spectral properties of the emitted field, we study the time evolution of two-point time correlation function for field operators, $G(z,\tau_1,\tau_2)$, given as 
\begin{equation}
 \begin{aligned}
  \frac{\partial G(z,\tau_1,\tau_2)}{\partial z} &= - \kappa(z) G(z,\tau_1,\tau_2) \\
  &+ \underbrace{\frac{3 \Delta o}{8 \pi} \Gamma_{sp} n_1 \left( \int\limits_0^{\tau_1} d\tau_1' e^{- \int\limits_{\tau_1'}^{\tau_1} d\tau'' \frac{\Gamma(z,\tau'')}{2}} \sigma(z,\tau_1') G(z,\tau_1',\tau_2) + \int\limits_0^{\tau_2} d\tau_2' e^{- \int\limits_{\tau_2'}^{\tau_2} d\tau'' \frac{\Gamma(z,\tau'')}{2}} \sigma(z,\tau_2') G(z,\tau_1,\tau_2')\right)}_{\text{Stimulated emission}} \\
  &+ \underbrace{\frac{3 \Delta o}{32 \pi \lambda^2} \Gamma_{sp} n_1 \int\limits_0^{\min(\tau_1,\tau_2)} d\tau' 
  e^{- \int\limits_{\tau'}^{\tau_1} d\tau'' \frac{\Gamma(z,\tau'')}{2}} e^{- \int\limits_{\tau'}^{\tau_2} d\tau'' \frac{\Gamma(z,\tau'')}{2}}
  \left( \gamma_{aug} b_{e} \rho_c(z,\tau') + J(z,\tau') \sigma_{abs} \rho_{ee}(z,\tau') \right)}_{\text{Spontaneous emission}},
 \end{aligned}
\end{equation}
where
\begin{equation}
 \Gamma(z,\tau) = \Gamma_{sp} + 2 \sigma_{abs} J(z,\tau) 
\end{equation}
is the total decay rate. The spectral properties can then be computed as
\begin{equation}
 \mathcal{I}(z,\omega) = \int d\tau_1 \; e^{i\omega \tau_1} \int d\tau_2 \; e^{-i\omega \tau_2} G(z,\tau_1,\tau_2).
\end{equation}

\subsection{Simulation results}

We present here details of the model calculations shown in the main text, for a pump photon energy of 73 eV. A reasonable agreement between the simulations according to equations presented above and the experimental dependence of emitted photon number on FEL pump energy and Xe gas pressure was found for the following simulation parameters: spontaneous lifetime corresponding to the transition from \ket{e} to \ket{g} level is 1 ns; Auger branching ratios are $b_e=$0.062/3 and  $b_g=$0.030/4, where the measured branching ratios (see Table \ref{tab:ratios}) are divided by degeneracy of the levels; XFEL absorption cross section $\sigma_{abs}$=5.2 Mb, superfluorescence absorption cross section $\sigma_{absF}$=60 Mb,  pump duration is 80 fs, Gaussian shape is assumed; $z_R=$2.0 mm, effective photon flux varies from $8.7\times 10^{8}$ photons/$\mu$m$^2$ at 30 mJ to $2.0\times 10^{9}$ photons/$\mu$m$^2$ at 70 mJ. The sample length is 6.5 mm.% beam waist diameter is 61 $\mu$m; sample length is 6.5 mm; the radiation is assumed to exit through an aperture with diameter 40 $\mu$m. (PLEASE CLARIFY THE APERTURE, SIZE OF THE BEAM...perhaps giving an effective beam radius is better?)\\

%The dependence of emitted radiation line width on the number of emitted photons presented in Fig. 4 of the main text was measured in a slightly different experimental condition. 3 consecutive pulses separated by 1 $\mu$s each were applied. In this case, the effective pressure in the cell is lowered, because the shock-wave induced by a pulse lowers the density at the interaction region, so that the subsequent pulse sees a rarefied medium. To account for this rather complex effect, we reduced the sample length to 3~mm. We have used the following parameters for calculations: branching ratios are $b_e=$0.044/3 and $b_g=$0.040/5; $z_R=$5.0 mm, beam waist size is 34 $\mu$m. (ALSO HERE PLEASE CLARIFY THE APERTURE, SIZE OF THE BEAM...perhaps giving an effective beam radius is better?) To determine the line broadening the simulation results were convoluted with Gaussian instrumental function having line width 0.415 meV.

\begin{figure}[htb]
\includegraphics[width=0.95\linewidth, scale=1]{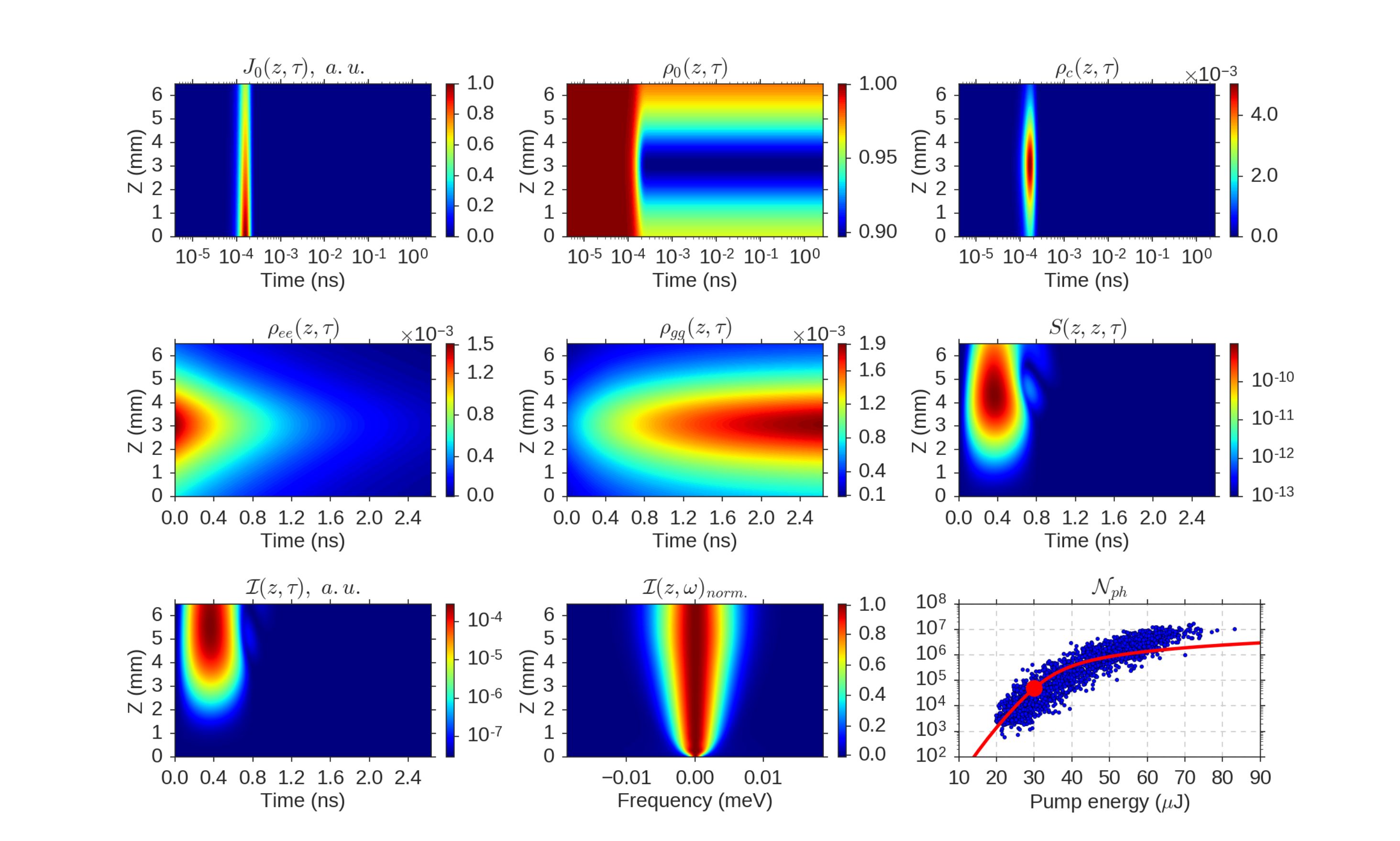}%
\caption{\label{fig: example_30mJ} Example of the evolution of the system observables in the case of 30~$\mu$J pump pulse energy and 73 eV pump photon energy. The number of emitted photons at this pump energy is shown by a red dot. The emitted radiation spectra for each $z$ value are normalized to maximal value at given $z$.}
\end{figure}

\begin{figure}[htb]
\includegraphics[width=0.95\linewidth, scale=1]{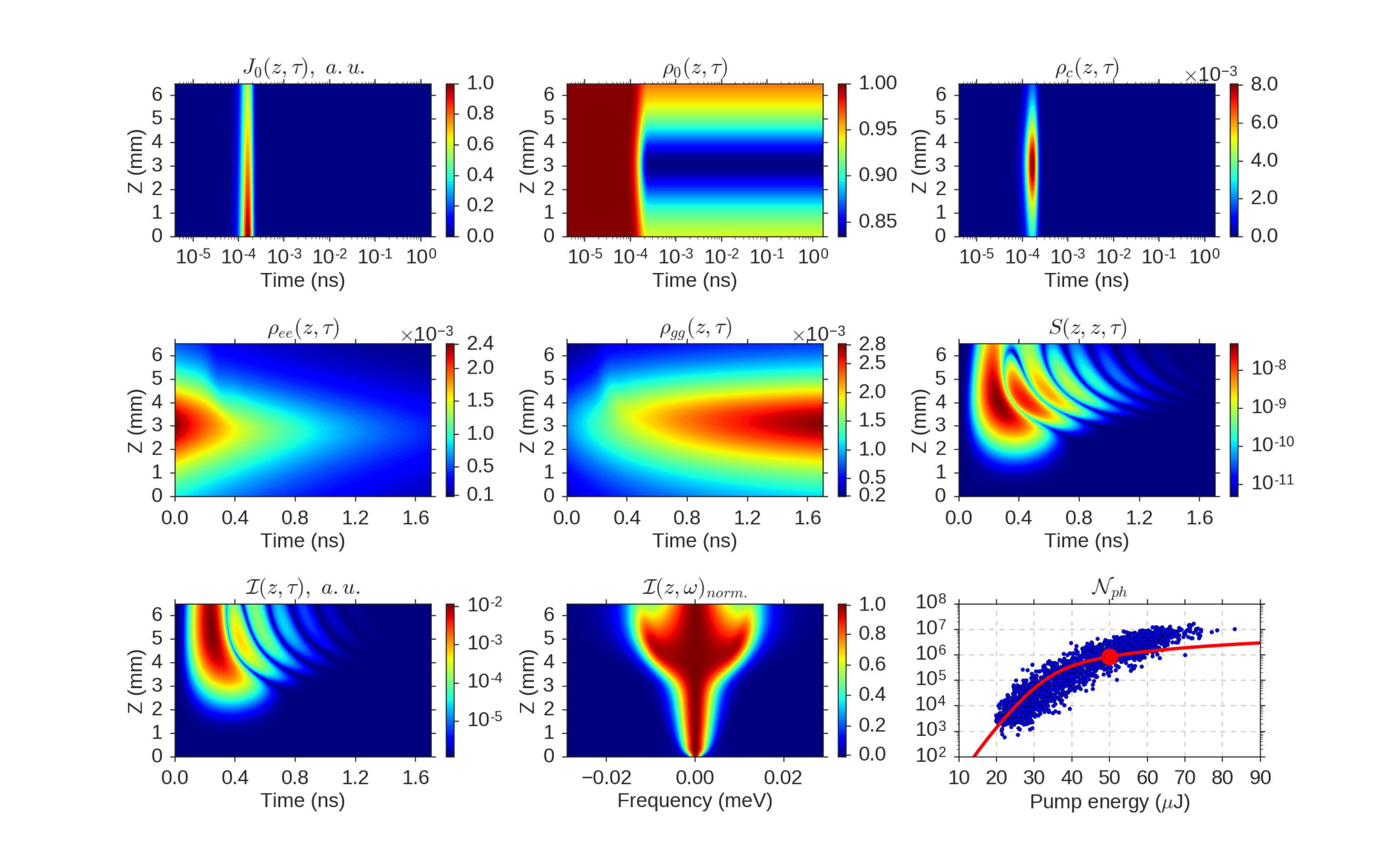}%
\caption{\label{fig: example_50mJ} Same as Fig. \ref{fig: example_30mJ} for 50$\mu$J pumping.}
\end{figure}

\begin{figure}[htb]
\includegraphics[width=0.95\linewidth, scale=1]{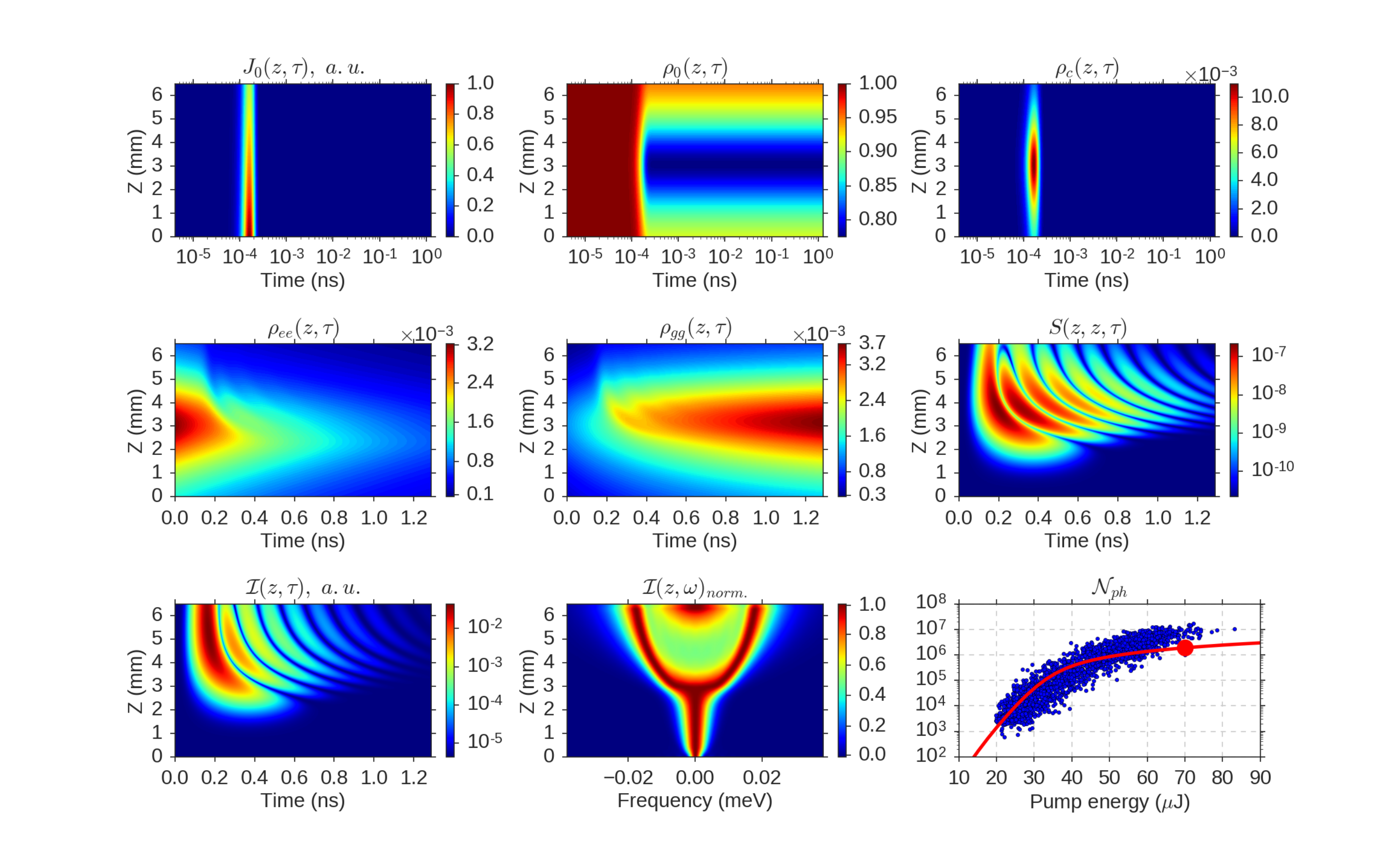}%
\caption{\label{fig: example_70mJ} Same as Fig. \ref{fig: example_30mJ} for 70$\mu$J pumping.}
\end{figure}

Figures \ref{fig: example_30mJ} - \ref{fig: example_70mJ} present examples of system observables evolution for various values of pumping. The same parameters as for Fig. 3 of the main text were used. Temporal and spectral intensity profiles at the exit of the medium for these simulations are shown in Figs. 4-b and 4-c of the main text. One can see that increasing of the pump results in stronger depletion of the ground level and higher population of the core-excited level. It is transferred to higher population of the \ket{e} level at the initial time. This in turn results in higher values of correlation of atomic coherences that leads to collective oscillatory behavior. The oscillatory behavior in the temporal domain translates in a broadening of the emitted radiation spectra. 

\clearpage

%\begin{figure}[htb]
%\includegraphics[width=1\linewidth, scale=1]{S10_73_65_pump_profiles_30_50_70uJ.pdf}%
%\caption{\label{fig: example_section_30_50_70mJ} Temporal and spectral intensity profiles at the exit of %the medium for cases presented on Figs.~\ref{fig: example_30mJ}~-~\ref{fig: example_70mJ}.}
%\end{figure} 
%\clearpage

%\bibliography{supplementary_biblio}